\documentclass[preprint]{aastex}
\usepackage{graphicx}
\usepackage{natbib}
\begin{document}

\title{Non-circular features in Saturn's D ring: D68}
\author{M.M. Hedman$^{a,*}$, J.A. Burt$^{b}$, J.A. Burns$^{c,d}$, M.R. Showalter$^e$}
\affil{\it  $^a$ Department of Physics, University of Idaho, Moscow ID 83843 USA \\
$^b$ Department of Astronomy and Astrophysics, University of California Santa Cruz, Santa Cruz, CA 95064 USA \\
$^c$ Department of Astronomy, Cornell University, Ithaca NY 14853 USA \\
$^d$ Department of Mechanical Engineering, Cornell University, Ithaca NY 14853 USA \\
$^e$ SETI Institute, Mountain View CA 94043 USA \\
${^*}$ Corresponding Author {\tt mhedman@uidaho.edu}}

\maketitle

\noindent {\bf Abstract:} D68 is a narrow ringlet located only 67,627 km (1.12 planetary radii) from Saturn's spin axis. Images of this ringlet obtained by the Cassini spacecraft reveal that this ringlet exhibits persistent longitudinal brightness variations and a substantial eccentricity ($ae=25\pm1$ km). By comparing observations made at different times, we confirm that the brightness variations revolve around the planet at approximately the local orbital rate ($1751.6^\circ$/day), and that the ringlet's pericenter precesses at $38.243^\pm0.008^\circ$/day, consistent with the expected apsidal precession rate at this location due to Saturn's higher-order gravitational harmonics. Surprisingly, we also find that the ringlet's semi-major axis appears to be decreasing with time at a rate of $2.4\pm0.4$ km/year between 2005 and 2013. A closer look at these measurements, along with a consideration of earlier Voyager observations of this same ringlet, suggests that the mean radius of D68 moves back and forth, perhaps with a period of around 15 Earth years or about half a Saturn year. These observations could place important constraints on both the ringlet's local dynamical environment and the planet's gravitational field.

{\bf Keywords:} Celestial Mechanics; Circumplanetary Dust; Planetary Rings; Saturn, Rings

\pagebreak

\section{Introduction}

Situated only 67,627 km (1.12 planetary radii) from Saturn's spin axis, D68 is the innermost discrete ringlet in Saturn's ring system. This ringlet was first seen in two images from the Voyager spacecraft \citep{Showalter96}, and has also been observed  multiple times by the cameras onboard the Cassini spacecraft \citep{Hedman07}. These data show that this ringlet is brightest when viewed at high phase angles, indicating that its optical properties are dominated by small particles less than 10 microns across. 

This ringlet is especially interesting because it is not azimuthally symmetric, but instead exhibits longitudinal variations in both its brightness and its radial position. Variations in the ringlet's radial position were first identified in images obtained early in the Cassini mission \citep{Hedman07}. However, some of these initial estimates of the ringlet's position were inaccurate and gave a misleading and confusing picture of the ringlet's shape. We have now re-examined the structure of this ring using more consistent image navigation procedures and a more extensive suite of observations spanning eight years. This new analysis reveals that, contra \citet{Hedman07},  D68 does in fact have a substantial eccentricity. Furthermore, we now detect longitudinal variations in the ringlet's brightness, as well as long-term evolution in its mean radial position. With this clearer picture of the ringlet's structure and dynamics, we are able to place better constraints on the forces perturbing the orbits of D68's particles, which include contributions from the higher-order components of Saturn's gravitational field.  

In Section~\ref{methods} below,  we describe the basic procedures used to extract estimates of the ringlet's position and integrated brightness from individual images.  The results of these calculations are presented in Section~\ref{results}. Here we first discuss several short observation sequences that illustrate both the longitudinal variations in the ringlet's brightness and its basic eccentric shape. Next, we compare measurements from a more extensive set of images in order to constrain the ringlet's precession rate and illustrate long-term variations in the ringlet's semi-major axis. Section~\ref{discussion} discusses  some implications of these measurements for the ringlet's dynamical environment, as well as constraints these observations could place on Saturn's gravitational field. Finally, Section~\ref{summary} recapitulates the primary findings of this analysis.

\section{Data Reduction Procedures}
\label{methods}

This investigation examines over 500 estimates of D68's brightness and/or radial position (distance from Saturn's center) derived from various images obtained by the Narrow Angle Camera (NAC) of the Imaging Science Subsystem onboard the Cassini spacecraft \citep{Porco04}. These brightness and position estimates were obtained by first calibrating and navigating each image,  converting the image brightness data into a radial brightness profile, and finally fitting the ringlet's profile to a Lorentzian peak function. Given the rather large number of images involved, the procedures we used to perform these calculations were designed to be as automatic as possible while still achieving adequately precise measurements of the ringlets' position and brightness. Position estimates accurate to within 1-2 km were considered sufficient for this analysis because the observed radial position of D68 varies by as much as 60 km, and the best D-ring images have resolutions of a few kilometers per pixel (see below). Also, this work only considers the brightness variations observed within several sequences of images covering a broad range of co-rotating ring longitudes in D68, so we were primarily concerned with obtaining accurate estimates of the ringlet's relative brightness in images obtained from roughly the same viewing geometry. 

All the relevant images were calibrated using the standard CISSCAL routines to convert raw data numbers into $I/F$, a standard measure of brightness that is unity for a Lambertian surface illuminated at normal incidence (see {\tt \small http://pds-rings.seti.org/cassini/iss/calibration.html}). 

Each image was initially navigated using the appropriate SPICE kernels to estimate to spacecraft's position and the camera's orientation relative to the rings. The camera's orientation was subsequently refined based on the positions of known stars in the field of view. Our previous analysis of D68 \citep{Hedman07} did not consistently use stars as a navigation reference, but instead employed stars for some images and the inner edge of the C ring for others, which led to some erroneous position estimates.  In particular, the radial locations of D68 in two images (N1493557225 and N1493559711) navigated using the C-ring were both off by 40 km (compare Table 1 in Hedman et al. 2007 with Table ~\ref{d68obs} below). Note the position of the C-ring's inner edge (defined as the location of maximum brightness slope) was mis-estimated by only about 5-10 km, which is much less than the error in D68's location. The origin of this discrepancy is still unclear, but since the C-ring edge and D68 were near opposite edges of these images, we suspect that there was an error in the images' effective scale. In principle, this could be due to a miscalculation of the spacecraft's range to the rings or the camera's distortion matrix, but both these options seem unlikely since other images taken around the same time did not exhibit such large geometry errors. Instead, we suspect that the earlier navigation included a slight twist and/or a shift in the azimuthal direction, both of which could shift the apparent position of D68 relative to the C ring edge. Regardless of how these shifts arose, we find that consistently using stars as a navigation reference yields a much more coherent set of position estimates for D68.

While taking any individual image, the camera pointed at a fixed location in the rings, so the spacecraft's motion caused the background stars to move through the field of view while the camera's shutter was open. Hence if the exposure was sufficiently long or the spacecraft was moving sufficiently fast, then the stars appeared as streaks of finite length rather than simple point sources. Depending on the length of these streaks, we used either a ``star-pointing'' or a ``streak-pointing'' procedure to navigate the images.

The ``star-pointing'' procedure was used whenever the streak length was less than twice the FWHM of the camera's point-spread function. In this case, the pixel coordinates $x,y$ of each star in the image are determined from a simple gaussian fit to the brightness data. These observed positions are then compared to the predicted star positions at the image mid-time (i.e., the time half-way through the exposure duration). The camera pointing is then adjusted to remove the average offsets between the observed and predicted coordinates of the relevant stars.  

For images where the stars form obviously extended streaks, we cannot simply fit the brightness data to a  gaussian, so we instead use a dedicated ``streak-pointing'' algorithm.  For each streak in an image, this program takes the streak's
brightness above background $I_*$ as a function of the pixel coordinates $x$ and $y$ and computes the coordinates $x_c, y_c$  of the streak's center-of-light:
\begin{equation}
x_c=\frac{\sum x I_*(x,y)}{\sum I_*(x,y)}, y_c=\frac{\sum y I_*(x,y)}{\sum I_*(x,y)},
\end{equation}
where the sums are over all pixels containing the streak. This center of light corresponds to the average star position in the image, and so the algorithm also computes the predicted position of each star at nine evenly spaced times between
the start and end of exposure duration, and averages these to generate a predicted center-of-light position. The camera pointing is then adjusted to remove any offsets between the observed and predicted center-of-light coordinates. 

Given the problems that previously arose from combining two different navigation methods, we have taken care to distinguish which algorithm was used to navigate each image below. Fortunately, it turns out that the two algorithms yield consistent position estimates for D68, which gives us some confidence in our revised navigation protocols. Note that both the above algorithms use multiple pixels to find each star's centroid or center-of-light, and so they can both potentially yield position estimates with the desired sub-pixel accuracy.   

After navigating each image, we construct a profile of brightness versus radius in Saturn's ring-plane by averaging over a range of longitudes. In most cases, we average over all longitudes visible in the image. However, we exclude regions in any image where the ring is obscured by Saturn's shadow or obvious stray-light artifacts \citep{West10}. For images obtained at low ring-opening angles, we only include longitudes close to the ring ansa where the radial resolution is highest.  

With the exception of the exceptionally high-resolution observations described by \citet{Hedman07}, D68 always appears in these profiles as a simple peak against a smooth background (see Section~\ref{origin}). Hence we fit the radial brightness profile in the vicinity to D68 to a Lorentzian plus linear background using the {\tt mpfitpeak} routine in IDL \citep{Markwardt09}. The parameters of this fit provide estimates of D68's location and integrated brightness in the image.\footnote{Only the profile from the image N1504582903 shows two distinct peaks, which are separated by about 15 km. We verified by eye that our algorithm found the location of the brighter peak in this image.}  The radial position of D68 is simply the peak location in the brightness profile, while the ringlet's integrated brightness is derived from the fitted peak amplitude $A$ and half-width at half-maximum $HWHM$. Specifically, we compute the ringlet's normal equivalent width or NEW as NEW=$\pi A \cdot HWHM \cdot \mu$, where $\mu$ is the cosine of the emission angle from the rings. NEW corresponds to the radially integrated brightness of the ringlet, and is a useful quantity for this analysis because it is insensitive to image resolution and geometry. These algorithms can also yield statistical uncertainties on the fit parameters, but these numbers are underestimates of the true uncertainties in the measurements because they neglect systematic effects due to image navigation, etc. Such systematic uncertainties are difficult to estimate {\it a priori}, so we will instead assess the errors in these measurements by considering the scatter of measurements within certain image sequences (see Section~\ref{results}).

\section{Results}
\label{results}

Using the above procedures, we were able to obtain position and/or brightness estimates from over 500 images of D68. These images include several ``movie sequences'' that provide the clearest evidence for longitudinal brightness variations in D68, as well as an observation sequence that imaged multiple longitudes nearly simultaneously and thus nicely illustrates D68's eccentricity. These data sets are described below in Sections~\ref{movies} and~\ref{retar}, respectively. Section~\ref{full} then discusses a larger suite of observations covering eight years and the constraints they place on D68's precession rate, while Section~\ref{drift} presents the evidence in these data for slow changes in the average radius of the ringlet.

\subsection{Longitudinal Brightness Variations from Movie Sequences}
\label{movies}

\begin{table}
\caption{Selected D68 movie sequences}
\label{d68mov}
\resizebox{6in}{!}{\begin{tabular}{|c|c|c|c|c|c|c|c|c|c|c|}\hline
Observation$^a$ & Navigation &Images & No.$^b$ & UTC & Duration & Inertial & Phase & Emission & Sub-Solar &   Radial  \\
& Method &   & & Time & &  Longitude$^c$ & Angle & Angle &  Longitude & Resolution \\ \hline
Rev 037 AZDKMRHP & star & N1547138243-N1547168273 &  73 & 2007-010T20:07:56 & 8.34 hours & 296.4$^\circ$ & 162.5$^\circ$ &  62.7$^\circ$ & 194.3$^\circ$ & 10.2 km \\
Rev 039 HIPHAMOVD & streak & N1550157836-N1550176479 &  60 & 2007-045T17:26:01 & 5.18 hours & 298.4$^\circ$ & 162.1$^\circ$ &  64.1$^\circ$ & 195.5$^\circ$ &  8.8 km \\
Rev 166 FNTLPMOV & star$^{d,e}$ & N1716447425-N1716468766 &  52 & 2012-144T09:03:11 & 5.93 hours & 205.4$^\circ$ &  42.9$^\circ$ &  92.8$^\circ$ & 255.2$^\circ$ & 10.0 km \\
Rev 168 DRCLOSE & streak & N1719551308-N1719564388 &  25 & 2012-180T06:05:11 & 3.63 hours & 148.7$^\circ$ & 148.4$^\circ$ & 106.9$^\circ$ & 256.3$^\circ$ &  4.2 km \\
Rev 173 DRNGMOV & star & N1728999806-N1729023557 &  30 & 2012-289T16:08:05 & 6.60 hours & 314.6$^\circ$ & 143.4$^\circ$ & 125.2$^\circ$ & 259.7$^\circ$ &  8.5 km \\
Rev 177 DRLPMOV & star & N1735244945-N1735264045 &  26 & 2012-361T22:14:19 & 5.31 hours & 202.5$^\circ$ &  41.7$^\circ$ & 101.6$^\circ$ & 261.9$^\circ$ &  8.6 km \\
Rev 180 DRLPMOV & star & N1738711767-N1738736572 &  56 & 2013-036T02:01:52 & 6.89 hours & 204.6$^\circ$ &  44.2$^\circ$ & 104.1$^\circ$ & 263.2$^\circ$ &  8.9 km \\
Rev 193 DRLPMOV & star & N1751146906-N1751160661 &  22 & 2013-179T22:40:46 & 3.82 hours & 192.7$^\circ$ &  30.0$^\circ$ &  95.1$^\circ$ & 267.7$^\circ$ &  7.5 km \\
Rev 198 DRNGMOV$^f$ & star & N1761014449-N1761035549 &  26 & 2013-294T04:39:59 & 5.86 hours & 316.0$^\circ$ & 139.3$^\circ$ & 132.4$^\circ$ & 271.3$^\circ$ & 14.0 km \\
Rev 198 DRNGMOV$^f$ & star & N1761035929-N1761057029 &  26 & 2013-294T10:37:59 & 5.86 hours & 166.0$^\circ$ & 138.1$^\circ$ & 131.1$^\circ$ & 271.3$^\circ$ & 15.1 km \\
Rev 199 DRNGMOV & star & N1765071135-N1765102855 &  62 & 2013-341T04:59:30 & 8.81 hours & 158.0$^\circ$ & 147.3$^\circ$ & 126.3$^\circ$ & 272.8$^\circ$ & 11.1 km \\
\hline
\end{tabular}}

$^a$ Note "Rev" refers to a Cassini orbit around Saturn

$^b$ Number of Images used to construct these profiles. Note that we do not necessarily use all the images in each sequence. Many movie sequences include images with different resolutions and/or exposures, and these different image properties could potentially complicate comparisons among the images within each sequence. We therefore only consider the set of images that yield the best signal-to-noise and resolution on D68. As shown in Figures~\ref{d68intcomp} and~\ref{d68poscomp}, some of the position and brightness estimates derived from these images are corrupted by stars or cosmic rays. Such questionable images are not included in subsequent analyses, which is why the numbers of images used in Tables~\ref{d68obs} and~\ref{d68obs2} sometimes deviates from the numbers given in this table.

$^c$ Longitude in Saturn's equatorial plane, relative to the intersection between that plane and Earth's equatorial plane in J2000 coordinates 

$^d$ Two of 52 images navigated using ring edges due to lack of stars

$^e$ Not used in model fitting due to low ring opening angles.

$^f$ This sequence observed two different locations in D68, and so we provide the geometry parameters for both parts of this observation separately.

\end{table}

\begin{figure}
\centerline{\resizebox{3.5in}{!}{\includegraphics{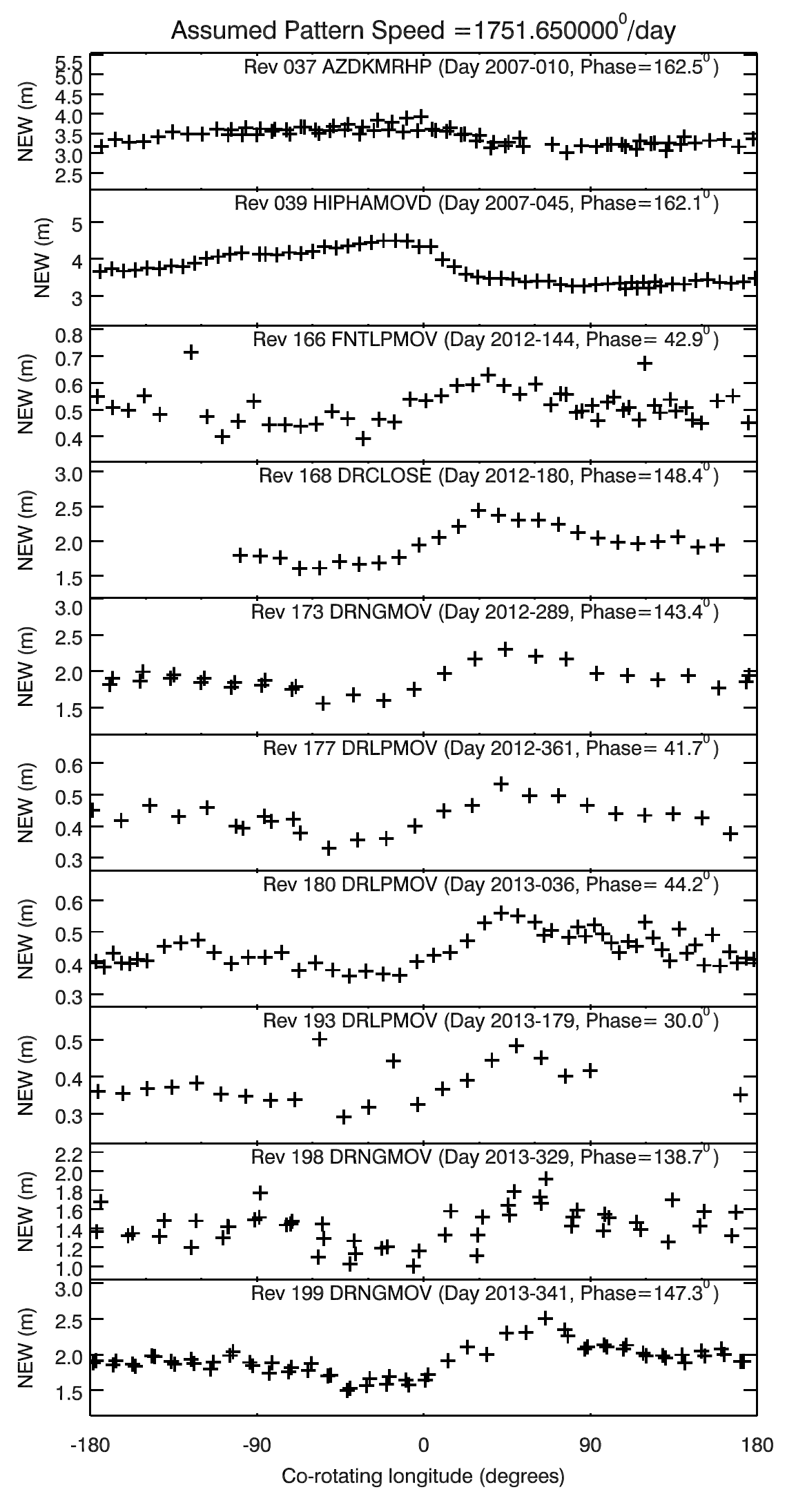}}}
\caption{Plots of D68's integrated brightness as a function of co-rotating longitude assuming a rotation rate of 1751.65$^\circ$/day and an epoch time of 2009-185T17:18:54UTC (300,000,000 TBD) derived from ten different movie sequences.  For the sake of clarity, error bars are not plotted on the data points, but are typically of order 10\% (the outliers in the Rev 166 data at around -120$^\circ$ and the Rev 193 data at -60$^\circ$ and -15$^\circ$ are due to corrupting stars or background features). The high-phase brightness data have all been corrected for phase-angle variations assuming the NEW is inversely proportional to the square of the scattering angle (which provides a good rough empirical match to the observed phase function). Note the brightness of the ring exhibits smooth variations with longitude, in particular a relatively steep slope  around  $0^\circ$.}
\label{d68intcomp}
\end{figure}

\begin{figure}
\resizebox{6in}{!}{\includegraphics{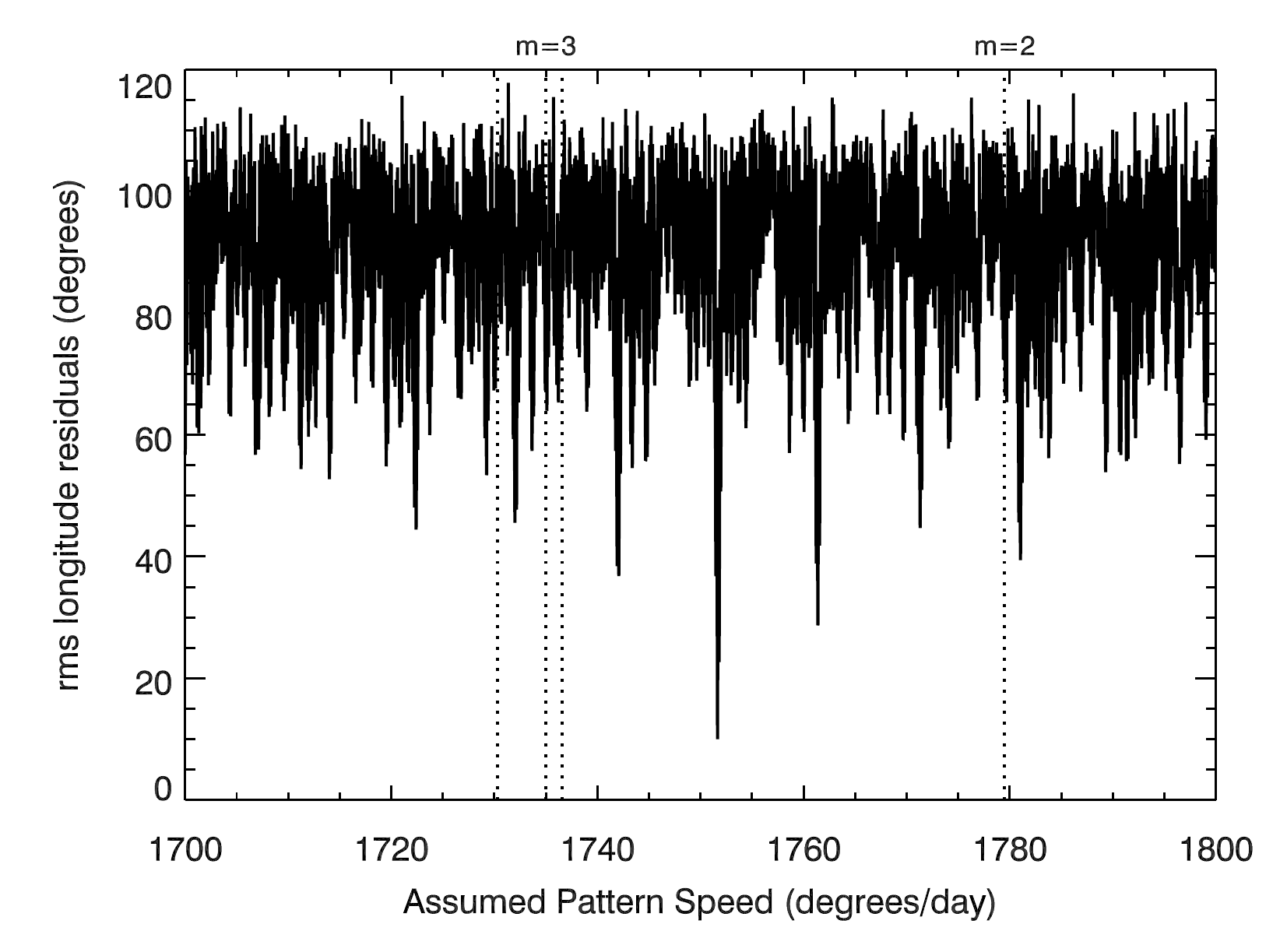}}
\caption{Plot comparing possible pattern speeds for D68's brightness variations. The curve shows the rms residuals of the brightness variations' co-rotating longitudes from a model where the longitudinal brightness variations move around the planet at a constant rate. The deepest minimum occurs at 1751.65$^\circ$/day, which is very close to the ringlet's predicted mean motion. Additional minima in the rms residuals probably represent aliasing of the sparse observations. The vertical lines mark the pattern speeds of the closest normal-mode oscillations identified by \citet{HN13}. Note that these resonant pattern speeds avoid all potential minima in the rms residuals.}
\label{d68mark}
\end{figure}

During a movie sequence, Cassini's camera points at a fixed inertial longitude in the rings and watches material rotate through the field of view for a time comparable to the local orbital period. Cassini obtained ten such image sequences of the region around D68 with resolutions and signal-to-noise sufficient to clearly detect that ringlet. Table~\ref{d68mov} provides the observation times and geometries for these movies. Note that two movies were obtained in early 2007, while the rest occurred in 2012-2013. 

There were no obvious localized structures in any individual image of D68. However, the average NEW of the ringlet varied systematically  by $\sim$40\% over the course of an orbital period. Figure~\ref{d68intcomp} illustrates these variations in a co-rotating longitude system with an orbital mean motion of 1751.65$^\circ$/day and an epoch time of 2009-185T17:18:54UTC (300,000,000 TDB), which is close to the expected mean motion of particles in this region (see below). 

All the observations obtained in 2012 or 2013 exhibit a clear rise in brightness with co-rotating longitude around 0$^\circ$, which leads to a broad peak around $+$45$^\circ$. Movies obtained at high and low phase angle display comparable fractional variations. Since particles of different sizes would contribute differently to the ringlet's brightness at different phase angles, this would imply that whatever process is responsible for making this feature is not sorting particles by size. Also, we see no obvious evolution in the brightness variations over time between 2012 and 2013. 

In contrast to the 2012-2013 observations, the most obvious feature in the 2007 movies is a clear decline in brightness near 0$^\circ$, so the brightness peak in these data is at small negative longitudes. This change in the shape of the profiles between 2007 and 2012 suggests that the brightness variations are evolving slowly over time.

The pattern speed at which these brightness variations revolve around the planet can be constrained by comparing the times and longitudes where a given feature was observed in the relevant movie sequences. In this case we determined the location of  the relatively sharp brightness shift around $0^\circ$ longitude in Figure~\ref{d68intcomp} as the longitude and time in each sequence where the slope of the brightness variations was largest.  From these coordinates, we then computed the corresponding co-rotating longitudes of this slope using a range of different possible pattern speeds in the vicinity of the ringlet's expected mean motion. The rms variations in those co-rotating longitudes should be at a minimum when the assumed pattern speed matches the true speed of the brightness variations. As shown in Figure~\ref{d68mark}, the deepest minimum occurs at  1751.65$^\circ$/day (additional minima arise due to aliasing of the sparse observations). Note that this pattern of minima is robust against whether we consider only the 2012-2013 data or if we include the 2007 data as well, so the changes in the pattern's morphology do not appear to have a dramatic effect on its observed speed. These brightness variations therefore seem to be rotating around the planet at roughly the same rate as the individual particles' mean motion.

\subsection{D68's shape from short-term observations}
\label{retar}

\begin{figure}
\centerline{\resizebox{5.5in}{!}{\includegraphics{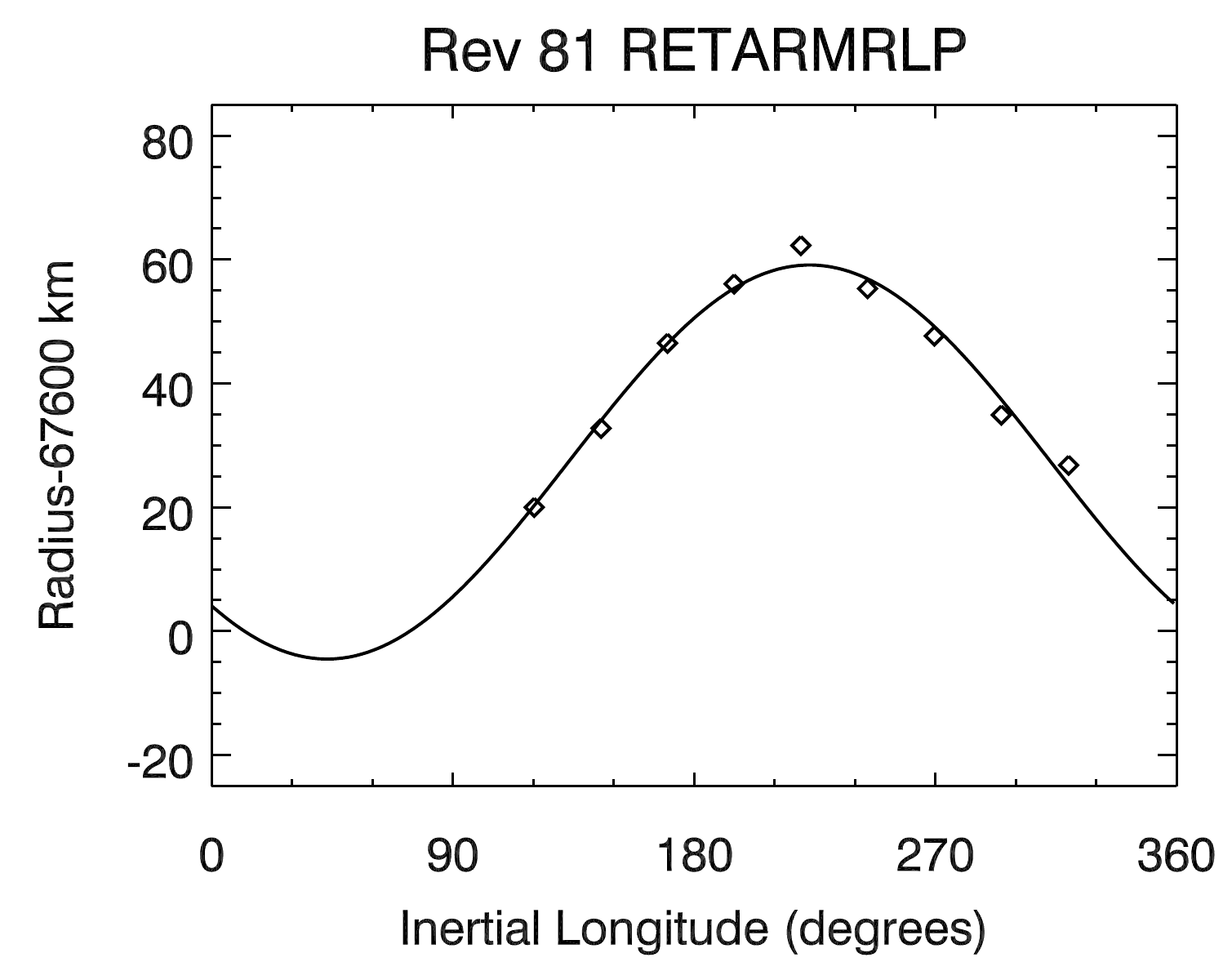}}}
\caption{D68's radial location as a function of inertial longitude derived in the Rev 81 RETARMRLP sequence, which observed the ringlet at multiple longitudes at almost the same time. Error-bars are not displayed, but the spatial resolution of these images are around 8 km, and the ringlet center can be determined to within 1-2 km in each image. The variations in the ringlet's radial position with longitude are clear and are consistent with those expected for an eccentric ringlet.}
\label{d68pos81ret}
\end{figure}

\begin{figure}
\centerline{\resizebox{3.5in}{!}{\includegraphics{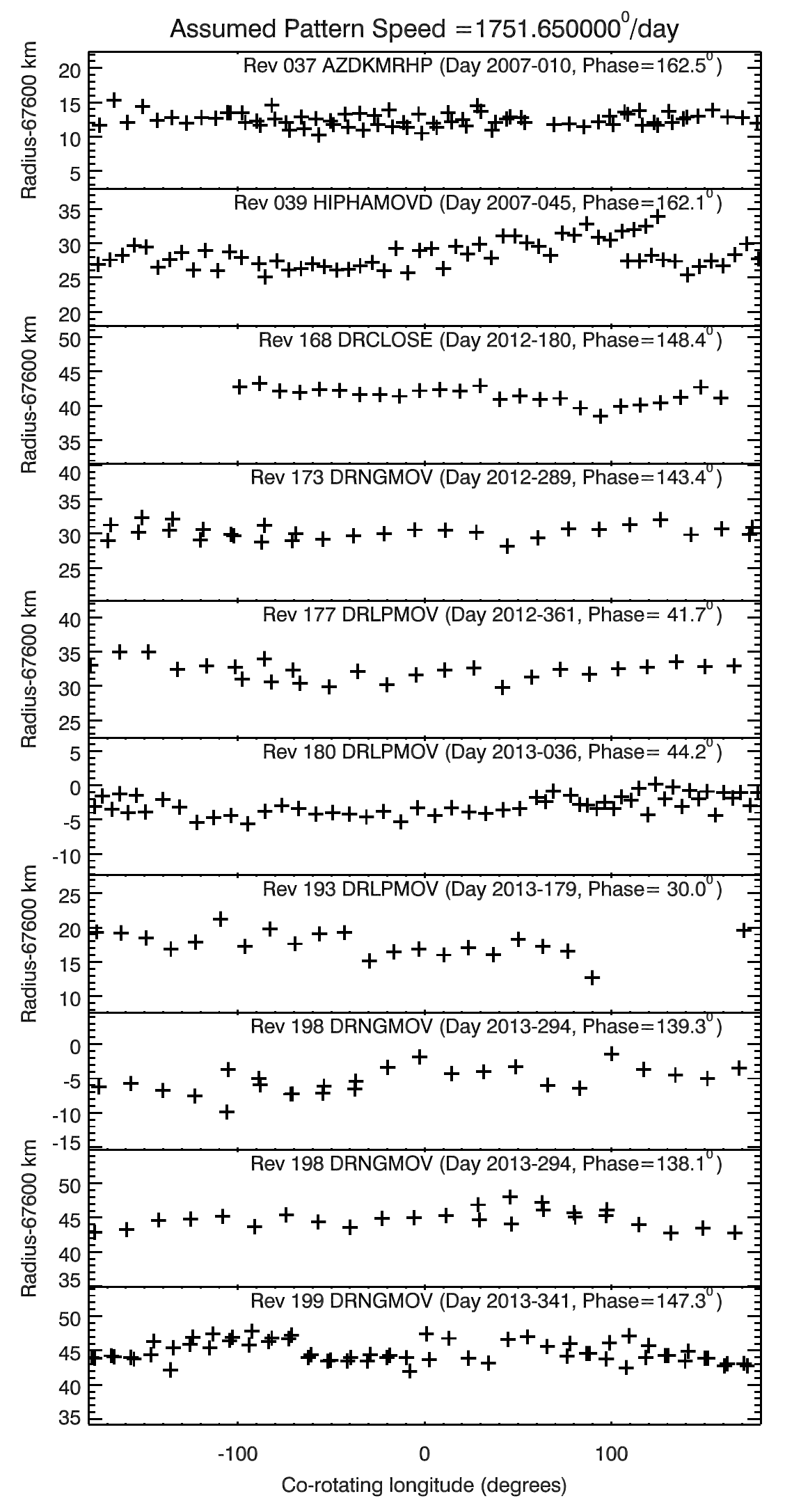}}}
\caption{Plots of D68's radial location as a function of co-rotating longitude (assuming an orbital rate of 1751.65$^\circ$/day and an epoch time of 2009-185T17:18:54UTC or 300,000,000 TDB) derived from nine different movie sequences (the Rev 166 FNTLPMOV data are not plotted because they could not be navigated as accurately, and the Rev 198 DRNGMOV data is split into two sets corresponding to measurements of two different longitudes).  For each movie, the radial position of the ringlet varies by only a few km, which is much less than the relevant image resolution.}
\label{d68poscomp}
\end{figure}

D68's basic shape is most clearly illustrated by an observation sequence called RETARMRLP obtained during Rev 81 in late 2008 (``Rev" designates Cassini's orbit around Saturn). Rather than stare at one point in the ring, this sequence imaged nine different inertial longitudes in 22 minutes (see data for images N1597747670-N1597748983 in Table~\ref{d68obs2}). Figure~\ref{d68pos81ret} shows D68's radial positions in these images versus inertial longitude $\lambda$. The ringlet's observed radial position varies by over 40 km among these images. Furthermore, the observed positions $r$ are consistent with the expected trends for an eccentric ringlet:
\begin{equation}
r=a-ae\cos(\lambda-\varpi),
\end{equation}
where $a, e$ and $\varpi$ are the ringlet's semi-major axis, eccentricity and pericenter longitude. To match these observations $ae$ must be between 25 and 30 km, which is consistent with fits to the full suite of position estimates obtained over the course of the Cassini mission to date (see Section~\ref{full} below).  This finding contradicts the claim made in \citet{Hedman07} that the ringlet cannot have a simple eccentric form, which was based on inaccurate position estimates for images N149355725 and N1493559711 (see Section~\ref{methods} above).

While the Rev 81 RETARMRLP sequence nicely illustrates D68's eccentricity, the 
various movie sequences described above can also help constrain the ringlet's shape. In these sequences, the camera stares at a fixed inertial longitude in the ring, and observes different particles as the ring material rotates through the field of view. Figure~\ref{d68poscomp} shows the ringlet's radial positions as functions of co-rotating longitude (Note the Rev 166 FNTLPMOV position estimates are not included here because these images did not always contain sufficient stars for proper navigation). For each movie, the ringlet's position is remarkably constant, indicating that the ringlet has a fixed or very slowly evolving shape (see below). Furthermore, the $rms$ dispersion of the position estimates for each movie is of order 1 or 2 km, which is an order of magnitude finer than the image resolution. This gives us some confidence that our navigation protocols do yield reliable and repeatable position estimates. 

More importantly, the data from these movie sequence constrain other aspects of D68's shape. In particular, since the ringlet's position at a given inertial longitude does not change by more than a few kilometers, we can place limits on the magnitudes of any normal-mode oscillations in the ringlet's radial position. These modes arise from non-trivial alignments in the pericenter locations and times of pericenter passage among particles found throughout the rings, which give rise to $m$-fold symmetric patterns in the ringlet's radial position that rotate around the planet at a speed:
\begin{equation}
\Omega_p=\frac{(m-1)n+\dot{\varpi}}{m},
\end{equation}
where $n$ is the local mean motion and $\dot{\varpi}$ is the local precession rate \citep{French91}. The case  $m=1$ corresponds to a freely-precessing eccentric ringlet, for which the pattern speed $\Omega_p$ is just the relevant precession rate. Modes with $m \ne 1$ have been observed in other narrow rings (e.g. Uranus' $\gamma$ and $\delta$ rings exhibit $m=0$ and $m=2$ normal modes, respectively, see French {\em et al.} 1991), but such modes are not visible in D68. If $m\ne 1$, then the pattern speed $\Omega_p$ is comparable to the particles' mean motion, so  we would expect to observe significant variations in the radial position of D68 at a given inertial longitude over the course of an orbital period. Since such variations are not observed in any of the D68 movie sequences, none of these modes can have an amplitude larger than a few kilometers. These data therefore indicate that we can treat D68 as a simple eccentric ringlet, whose pericenter should precess slowly around the planet.

\subsection{Constraining D68's precession rate}
\label{full}

\begin{table}
\caption{Summary of D68 observations that were navigated with stars}
\label{d68obs}
\resizebox{6in}{!}{\begin{tabular}{|c|c|c|c|c|c|c|c|c|c|}\hline
Images & No.$^a$ & UTC & Inertial & Phase & Emission & Sub-Solar & D68 & Radial & rms \\
 & & Time & Longitude$^b$ & Angle & Angle &  Longitude & Radius & Resolution & Variations \\ \hline
N1493557225 &   1 & 2005-120T12:33:19 &  55.5$^\circ$ &  33.4$^\circ$ & 109.5$^\circ$ & 171.8$^\circ$ &  67609.6 km &  8.9 km & --- \\
N1493559711 &   1 & 2005-120T13:14:45 & 220.5$^\circ$ &  38.5$^\circ$ & 109.5$^\circ$ & 171.8$^\circ$ &  67657.9 km &  8.9 km & --- \\
N1496894416 &   1 & 2005-159T03:32:49 & 286.5$^\circ$ &  20.8$^\circ$ & 106.2$^\circ$ & 173.3$^\circ$ &  67651.0 km &  2.1 km & --- \\
N1504582863 &   1 & 2005-248T03:12:47 & 253.2$^\circ$ &  12.1$^\circ$ & 105.8$^\circ$ & 176.6$^\circ$ &  67612.9 km &  2.4 km & --- \\
N1504584256 &   1 & 2005-248T03:36:01 &  97.2$^\circ$ &  10.6$^\circ$ & 105.7$^\circ$ & 176.6$^\circ$ &  67649.9 km &  2.4 km & --- \\
N1541986518-N1541986794 &   2 & 2006-316T01:05:19 &  70.6$^\circ$- 77.7$^\circ$ & 150.4$^\circ$-150.6$^\circ$ &  46.9$^\circ$- 47.1$^\circ$ & 192.2$^\circ$ &  67647.8 km &  8.6 km- 8.8 km &   0.3 km \\
N1544384451 &   1 & 2006-343T19:08:16 & 291.5$^\circ$ & 161.7$^\circ$ &  69.6$^\circ$ & 193.2$^\circ$ &  67647.4 km & 10.4 km & --- \\
N1544384801 &   1 & 2006-343T19:14:06 & 312.2$^\circ$ & 161.6$^\circ$ &  69.4$^\circ$ & 193.2$^\circ$ &  67650.4 km & 10.3 km & --- \\
N1547138243-N1547168273 &  73 & 2007-010T20:14:50 & 296.0$^\circ$-296.0$^\circ$ & 162.2$^\circ$-162.7$^\circ$ &  61.4$^\circ$- 64.1$^\circ$ & 194.3$^\circ$ &  67612.7 km & 10.3 km-10.1 km &   0.5 km \\
N1549740228 &   1 & 2007-040T18:50:42 &  82.0$^\circ$ & 112.7$^\circ$ &  32.5$^\circ$ & 195.3$^\circ$ &  67617.5 km & 11.3 km & --- \\
N1551438946 &   1 & 2007-060T10:42:29 &  82.8$^\circ$ & 117.7$^\circ$ &  32.7$^\circ$ & 196.0$^\circ$ &  67609.2 km & 10.4 km & --- \\
N1552930916-N1552931147 &   2 & 2007-077T17:10:25 &  79.1$^\circ$- 83.3$^\circ$ & 114.6$^\circ$-114.8$^\circ$ &  35.1$^\circ$- 35.2$^\circ$ & 196.6$^\circ$ &  67626.7 km & 10.8 km-10.5 km &   1.0 km \\
N1575636127 &   1 & 2007-340T12:06:15 & 297.8$^\circ$ &  14.6$^\circ$ &  84.9$^\circ$ & 205.3$^\circ$ &  67612.0 km & 10.1 km & --- \\
N1619794259 &   1 & 2009-120T14:09:53 & 285.2$^\circ$ &  74.7$^\circ$ &  26.4$^\circ$ & 221.4$^\circ$ &  67646.5 km &  9.1 km & --- \\
N1627206524-N1627207409 &   7 & 2009-206T09:14:08 & 293.7$^\circ$-294.0$^\circ$ & 159.5$^\circ$-159.9$^\circ$ &  96.7$^\circ$- 96.9$^\circ$ & 224.0$^\circ$ &  67648.7 km &  5.6 km- 5.6 km &   0.3 km \\
N1628920479-N1628921881 &  10 & 2009-226T05:24:05 & 228.1$^\circ$-228.7$^\circ$ &  97.0$^\circ$- 97.4$^\circ$ &  73.3$^\circ$- 73.3$^\circ$ & 224.6$^\circ$ &  67612.5 km &  9.3 km- 9.4 km &   0.9 km \\
N1628942889-N1628943346 &   9 & 2009-226T11:29:43 & 229.7$^\circ$-229.8$^\circ$ &  98.7$^\circ$- 98.7$^\circ$ &  73.7$^\circ$- 73.7$^\circ$ & 224.7$^\circ$ &  67611.3 km &  9.8 km- 9.8 km &   0.7 km \\
N1628993140-N1628993597 &   9 & 2009-227T01:27:13 & 232.1$^\circ$-232.8$^\circ$ & 101.9$^\circ$-101.9$^\circ$ &  74.6$^\circ$- 74.6$^\circ$ & 224.7$^\circ$ &  67606.0 km & 10.7 km-10.7 km &   1.0 km \\
N1629005410-N1629005867 &   9 & 2009-227T04:51:43 & 231.9$^\circ$-232.1$^\circ$ & 102.6$^\circ$-102.6$^\circ$ &  74.8$^\circ$- 74.8$^\circ$ & 224.7$^\circ$ &  67606.7 km & 10.9 km-11.0 km &   2.0 km \\
N1629139233-N1629139698 &   9 & 2009-228T18:02:11 & 239.5$^\circ$-242.5$^\circ$ & 109.2$^\circ$-109.2$^\circ$ &  76.7$^\circ$- 76.7$^\circ$ & 224.7$^\circ$ &  67603.7 km & 12.6 km-12.6 km &   2.1 km \\
N1629151495-N1629151968 &   9 & 2009-228T21:26:39 & 242.5$^\circ$-242.5$^\circ$ & 109.7$^\circ$-109.7$^\circ$ &  76.8$^\circ$- 76.8$^\circ$ & 224.7$^\circ$ &  67602.3 km & 12.7 km-12.7 km &   1.5 km \\
N1629423594-N1629423890 &   6 & 2009-232T01:00:08 & 255.0$^\circ$-255.0$^\circ$ & 120.3$^\circ$-120.3$^\circ$ &  80.1$^\circ$- 80.1$^\circ$ & 224.8$^\circ$ &  67642.1 km & 13.7 km-13.6 km &   1.8 km \\
N1629442347-N1629442640 &   6 & 2009-232T06:12:39 & 255.0$^\circ$-255.0$^\circ$ & 121.0$^\circ$-121.0$^\circ$ &  80.4$^\circ$- 80.4$^\circ$ & 224.8$^\circ$ &  67646.3 km & 13.6 km-13.6 km &   2.7 km \\
N1629509454-N1629509630 &   4 & 2009-233T00:50:07 & 260.0$^\circ$-260.0$^\circ$ & 123.7$^\circ$-123.7$^\circ$ &  81.2$^\circ$- 81.2$^\circ$ & 224.9$^\circ$ &  67647.7 km & 13.3 km-13.3 km &   1.1 km \\
N1629527120-N1629528380 &   8 & 2009-233T05:53:35 & 260.0$^\circ$-260.0$^\circ$ & 124.4$^\circ$-124.4$^\circ$ &  81.5$^\circ$- 81.5$^\circ$ & 224.9$^\circ$ &  67647.4 km & 13.2 km-13.2 km &   1.3 km \\
N1629545870-N1629547130 &   8 & 2009-233T11:06:05 & 260.0$^\circ$-260.0$^\circ$ & 125.2$^\circ$-125.2$^\circ$ &  81.7$^\circ$- 81.7$^\circ$ & 224.9$^\circ$ &  67648.8 km & 13.1 km-13.1 km &   1.8 km \\
N1630301204 &   1 & 2009-242T04:44:24 & 210.7$^\circ$ &  73.2$^\circ$ &  78.1$^\circ$ & 225.1$^\circ$ &  67646.3 km &  9.5 km & --- \\
N1632477236 &   1 & 2009-267T09:11:21 & 215.0$^\circ$ &  80.1$^\circ$ &  78.8$^\circ$ & 225.9$^\circ$ &  67623.7 km & 11.8 km & --- \\
N1719549271 &   1 & 2012-180T03:42:15 & 322.2$^\circ$ & 154.9$^\circ$ & 108.4$^\circ$ & 256.3$^\circ$ &  67594.2 km &  4.3 km & --- \\
N1728999806-N1729023557 &  30 & 2012-289T16:08:05 & 312.4$^\circ$-317.0$^\circ$ & 141.5$^\circ$-145.4$^\circ$ & 124.6$^\circ$-125.8$^\circ$ & 259.7$^\circ$ &  67630.2 km &  8.7 km- 8.2 km &   0.4 km \\
N1735244945-N1735264045 &  26 & 2012-361T22:14:19 & 202.5$^\circ$-202.5$^\circ$ &  40.3$^\circ$- 43.0$^\circ$ & 100.5$^\circ$-102.6$^\circ$ & 261.9$^\circ$ &  67632.2 km &  8.5 km- 8.7 km &   0.5 km \\
N1738711767-N1738736572 &  55 & 2013-036T02:01:52 & 203.5$^\circ$-205.7$^\circ$ &  42.5$^\circ$- 45.8$^\circ$ & 102.8$^\circ$-105.4$^\circ$ & 263.2$^\circ$ &  67597.3 km &  8.8 km- 9.1 km &   0.6 km \\
N1751148216-N1751160661 &  18 & 2013-179T22:51:41 & 192.7$^\circ$-192.8$^\circ$ &  28.9$^\circ$- 31.4$^\circ$ &  94.1$^\circ$- 96.3$^\circ$ & 267.7$^\circ$ &  67617.7 km &  7.5 km- 7.6 km &   0.5 km \\
N1761014449-N1761035549 &  26 & 2013-294T04:39:59 & 316.0$^\circ$-316.0$^\circ$ & 138.4$^\circ$-140.1$^\circ$ & 132.0$^\circ$-132.9$^\circ$ & 271.3$^\circ$ &  67594.8 km & 13.9 km-14.1 km &   1.1 km \\
N1761035929-N1761057029 &  26 & 2013-294T10:37:59 & 166.0$^\circ$-166.0$^\circ$ & 137.2$^\circ$-139.0$^\circ$ & 130.6$^\circ$-131.5$^\circ$ & 271.3$^\circ$ &  67645.2 km & 14.9 km-15.2 km &   0.8 km \\
N1765071135-N1765102855 &  62 & 2013-341T04:59:30 & 158.0$^\circ$-158.0$^\circ$ & 145.2$^\circ$-149.4$^\circ$ & 124.9$^\circ$-127.6$^\circ$ & 272.7$^\circ$ &  67644.8 km & 10.9 km-11.3 km &   0.6 km \\
\hline
\end{tabular}}

$^a$ Number of images.

$^b$ Longitude in Saturn's equatorial plane, relative to the intersection between that plane and Earth's equatorial plane in J2000 coordinates 
\end{table}

\begin{table}
\caption{Summary of D68 observations that were navigated with star streaks}
\label{d68obs2}
\resizebox{6in}{!}{\begin{tabular}{|c|c|c|c|c|c|c|c|c|c|}\hline
Images & No.$^a$ & UTC & Inertial & Phase & Emission & Sub-Solar & D68 & Radial & rms \\
 & & Time & Longitude$^b$ & Angle & Angle &  Longitude & Radius & Resolution & Variations \\ \hline
N1532676869 &   1 & 2006-208T07:03:02 & 248.0$^\circ$ & 150.4$^\circ$ &  75.8$^\circ$ & 188.5$^\circ$ &  67660.5 km & 10.8 km & --- \\
N1536743096 &   1 & 2006-255T08:33:03 &  81.3$^\circ$ & 161.0$^\circ$ &  70.0$^\circ$ & 190.1$^\circ$ &  67615.9 km &  9.1 km & --- \\
N1541397571 &   1 & 2006-309T05:27:07 & 307.3$^\circ$ & 154.4$^\circ$ &  77.5$^\circ$ & 192.0$^\circ$ &  67608.7 km &  9.4 km & --- \\
N1543426394 &   1 & 2006-332T17:00:34 & 309.9$^\circ$ & 157.8$^\circ$ &  74.6$^\circ$ & 192.8$^\circ$ &  67660.2 km &  9.8 km & --- \\
N1546070460 &   1 & 2006-363T07:28:14 & 133.0$^\circ$ & 131.8$^\circ$ & 106.6$^\circ$ & 193.9$^\circ$ &  67629.8 km &  6.1 km & --- \\
N1546071691 &   1 & 2006-363T07:48:45 & 147.6$^\circ$ & 131.6$^\circ$ & 106.7$^\circ$ & 193.9$^\circ$ &  67626.4 km &  6.1 km & --- \\
N1550157836-N1550176479 &  60 & 2007-045T17:26:02 & 297.4$^\circ$-299.3$^\circ$ & 161.9$^\circ$-162.2$^\circ$ &  62.7$^\circ$- 65.5$^\circ$ & 195.5$^\circ$ &  67628.7 km &  8.9 km- 8.7 km &   0.5 km \\
N1597747670 &   1 & 2008-231T10:09:13 & 317.4$^\circ$ &  65.3$^\circ$ &  30.2$^\circ$ & 213.5$^\circ$ &  67626.8 km &  7.8 km & --- \\
N1597747807 &   1 & 2008-231T10:11:30 & 293.1$^\circ$ &  67.7$^\circ$ &  27.9$^\circ$ & 213.5$^\circ$ &  67634.9 km &  7.6 km & --- \\
N1597747954 &   1 & 2008-231T10:13:57 & 269.1$^\circ$ &  70.1$^\circ$ &  25.5$^\circ$ & 213.5$^\circ$ &  67647.7 km &  7.6 km & --- \\
N1597748127 &   1 & 2008-231T10:16:50 & 245.1$^\circ$ &  71.9$^\circ$ &  23.7$^\circ$ & 213.5$^\circ$ &  67655.3 km &  7.8 km & --- \\
N1597748316 &   1 & 2008-231T10:19:59 & 221.1$^\circ$ &  72.9$^\circ$ &  23.0$^\circ$ & 213.5$^\circ$ &  67662.3 km &  7.9 km & --- \\
N1597748508 &   1 & 2008-231T10:23:11 & 196.9$^\circ$ &  72.8$^\circ$ &  23.9$^\circ$ & 213.5$^\circ$ &  67656.1 km &  7.7 km & --- \\
N1597748689 &   1 & 2008-231T10:26:12 & 172.5$^\circ$ &  71.7$^\circ$ &  25.7$^\circ$ & 213.5$^\circ$ &  67646.5 km &  7.5 km & --- \\
N1597748849 &   1 & 2008-231T10:28:52 & 147.7$^\circ$ &  69.9$^\circ$ &  28.1$^\circ$ & 213.5$^\circ$ &  67632.8 km &  7.5 km & --- \\
N1597748983 &   1 & 2008-231T10:31:06 & 122.3$^\circ$ &  67.6$^\circ$ &  30.3$^\circ$ & 213.5$^\circ$ &  67620.0 km &  7.9 km & --- \\
N1641835989 &   1 & 2010-010T16:49:26 & 325.2$^\circ$ & 160.2$^\circ$ & 111.3$^\circ$ & 229.2$^\circ$ &  67645.0 km &  2.8 km & --- \\
N1719550962-N1719565362 &  28 & 2012-180T06:10:26 & 144.0$^\circ$-152.5$^\circ$ & 144.8$^\circ$-152.6$^\circ$ & 106.0$^\circ$-107.8$^\circ$ & 256.3$^\circ$ &  67641.7 km &  4.4 km- 3.7 km &   0.3 km \\
N1721658014 &   1 & 2012-204T13:27:45 & 359.9$^\circ$ & 169.3$^\circ$ &  99.6$^\circ$ & 257.1$^\circ$ &  67640.9 km &  2.6 km & --- \\
N1729211907-N1729217787 &  14 & 2012-292T00:34:09 & 198.9$^\circ$-204.0$^\circ$ & 134.2$^\circ$-140.6$^\circ$ &  74.6$^\circ$- 78.7$^\circ$ & 259.8$^\circ$ &  67615.3 km &  3.2 km- 3.1 km &   0.3 km \\
N1729223727 &   1 & 2012-292T03:02:09 & 209.9$^\circ$ & 127.2$^\circ$ &  70.3$^\circ$ & 259.8$^\circ$ &  67614.6 km &  3.0 km & --- \\
\hline
\end{tabular}}

$^a$ Number of images.

$^b$ Longitude in Saturn's equatorial plane, relative to the intersection between that plane and Earth's equatorial plane in J2000 coordinates 
\end{table}

\begin{figure}
\resizebox{6in}{!}{\includegraphics{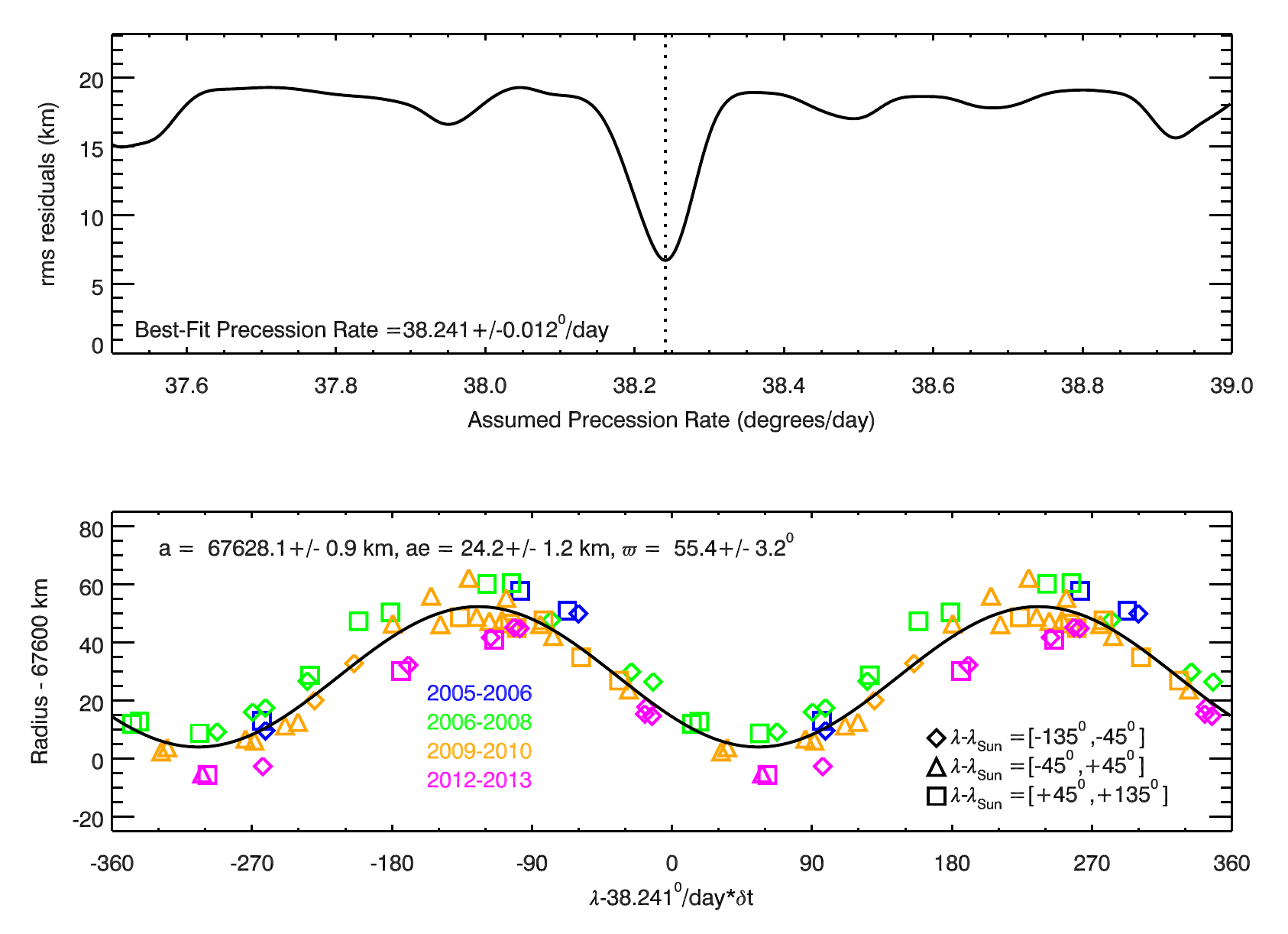}}
\caption{Fits of D68's radial position to a precessing-eccentric-ringlet model. The top panel shows the $rms$ of the fit-residuals as a function of the assumed pattern speed. The best-fit pattern speed occurs at 38.241$^\circ$/day, where the $rms$ variations are roughly 7 km. The bottom panel plots the observed data and the best fitting model. Different symbols indicate the observed longitude relative to the Sun and the colors designate the observation date.}
\label{d68fig}
\end{figure}

\begin{figure}
\resizebox{6in}{!}{\includegraphics{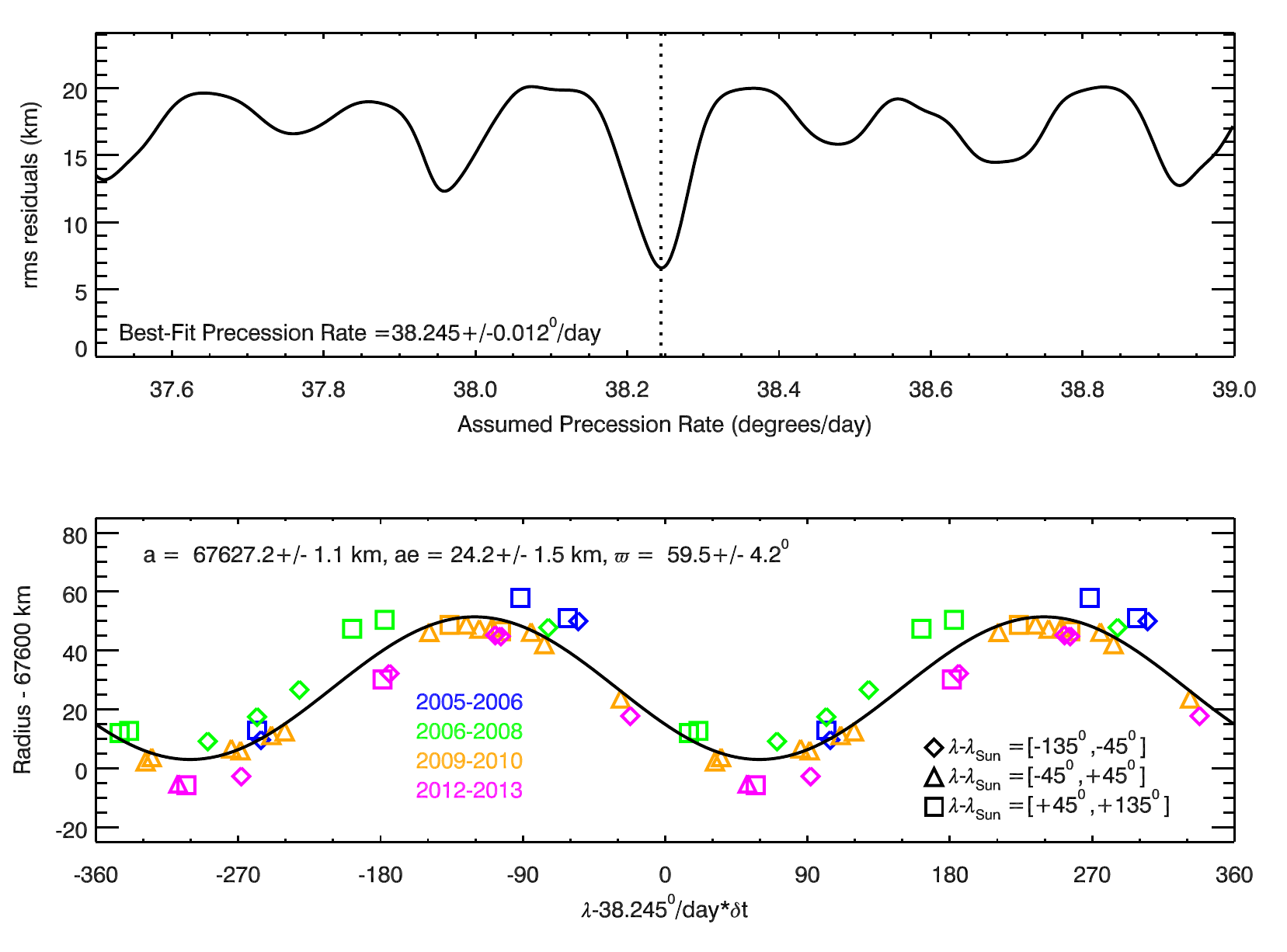}}
\caption{Fits of D68's radial position to a precessing-eccentric-ringlet model, using only data from images that could be navigated with the star-pointing algorithm. The top panel shows the $rms$ of the fit-residuals as a function of the assumed pattern speed. The best-fit pattern speed occurs at 38.245$^\circ$/day, where the $rms$ variations are roughly 7 km. The bottom panel plots the observed data and the best fitting model. Different symbols indicate the observed longitude relative to the Sun and the colors designate the observation date.}
\label{d68fig2}
\end{figure}

While individual image sequences illustrate D68's basic eccentric shape,  this ringlet should also steadily precess around Saturn due to the planet's finite oblateness. Hence we need a larger array of images covering a broad span of time in order to precisely determine the ringlet's eccentricity, semi-major axis and precession rate.  We therefore searched for images of D68 with sufficient signal-to-noise to detect the ringlet, resolutions better than 20 km/pixel and ring opening angles greater than about 5$^\circ$. These images included representatives from all of the movie sequences described above, except for the Rev 166 FNTLPMOV and Rev 193 DRNGMOV sequences, whose opening angles were below 5$^\circ$. We elected to include the Rev 193 DRNGMOV data in the analysis because they were one of the most recent D68 observations available, but we excluded the Rev 166 FNTLPMOV data because they were more difficult to navigate due to a lack of visible stars and the significantly lower ring opening angle. We also excluded various images (including a few from the movie sequences) where the profiles were obviously corrupted by stars or other imaging artifacts. This left 540 estimates of D68's radial position derived 420 star-pointed images and 120 streak-pointed images. (A table providing the observation name, image time, radial position estimates and viewing geometry for each image will be made available as an electronic supplement to this article.) 

We do not estimate the ringlet's shape or precession rate directly from the 540 individual position estimates, because these values include data from a number of movie sequences (both the long movies described above and shorter movies covering a smaller fraction of co-rotating longitudes). Since D68's pericenter precesses at a much slower rate ($\sim38.2^\circ$/day, see below) than the local orbital speed, these sequences provide many replicate measures of the ringlet's position at a single true anomaly. Because image processing errors might systematically bias these replicated measurements, we group together any image set where adjacent images are less than 10$^\circ$ apart in inertial longitude and less than 1 hour apart in time. For each of these groups, we derive a single position estimate as the mean of all individual estimates. Tables~\ref{d68obs} and~\ref{d68obs2} list these estimates, along with the relevant observation times and geometries. They also provide the resolution of the relevant images and the $rms$ variations in the radius estimates within each group. Note that these $rms$ variations are much less than the image resolution, confirming the robustness of our navigation estimates.

Based on the short observation sequences discussed above, we postulate that the ringlet has a simple eccentric shape, so that its radial position $r$ can be expressed as the following function of inertial longitude $\lambda$ and time $\delta t$ (measured relative to an epoch time $t_o=$ 300,000,000 TDB, which corresponds to 2009-185T17:18:54UTC):
\begin{equation}
r=a-ae\cos(\lambda-\dot{\varpi}\delta t-\varpi_o),
\label{rmod}
\end{equation} 
where  $a, e,  \dot{\varpi}$ and $\varpi_o$  are the ringlet's semi-major axis, eccentricity, precession rate and pericenter longitude at the epoch time. In practice we determine these parameters using a two-step procedure. First, we estimate the precession rate by finding the value of $\dot\varpi$ that minimizes the $rms$ residuals from the above model. Then, holding the precession rate at the best-fit value, we use a least-squares fit to determine the remaining orbital parameters. 

Based on the \citet{Jacobson06} model for Saturn's gravity field, the precession rate of D68 should be $\sim$38.2$^\circ$/day. Thus we consider $\dot\varpi$ values between 37$^\circ$/day and 39$^\circ$/day separated by 0.001$^\circ$/day for this analysis. For each value of $\dot\varpi$, we perform a least-squared fit of the data listed in Tables~\ref{d68obs} and~\ref{d68obs2} to the above model. All these data were given the same weight in the fit even though some radial position estimates came from a single image while others came from long movie sequences.\footnote{This choice can be justified by noting that the $rms$ dispersion around the best-fitting model is much larger than the $rms$ variations within any group, which indicates that statistical uncertainties are not the dominant source of error. Instead systematic effects such as navigational errors or un-modeled ring structure are probably responsible for much of the scatter in the position estimates, and these systematic errors are difficult to estimate {\it a priori}.}  The top panels of Figures~\ref{d68fig} and~\ref{d68fig2} show the $rms$ residuals from the best-fit model versus the assumed precession rate. Figure~\ref{d68fig} is derived from an analysis of all the data, while Figure~\ref{d68fig2} only includes data  that were navigated with the star-pointing algorithm. Both  curves show a clear minimum between 38.24$^\circ/$day and 38.25$^\circ$/day.  To estimate the uncertainty on these precession rates, we assume the minimum $rms$ value ($\sim$ 7 km) corresponds to the real statistical uncertainty in the position estimates, and use this number to compute the relevant $\chi^2$ statistic and the corresponding probability to exceed as a function of precession rate.  The probability statistic exhibits a sharp peak, whose gaussian width provides an estimate of the uncertainty in the precession rate of around 0.012$^\circ/$day. Note that this error is conservative, because the $rms$ scatter around the model includes systematic trends (see below).

Assuming the best-fit value for $\dot\varpi$, we then estimate the values for $a$, $ae$ and $\varpi_o$ using an unweighted least-squares fit. The uncertainties on these parameters are computed assuming all data points have an error equal to the $rms$ dispersion around the best-fit model (again, around 7 km). These parameter estimates and uncertainties are given in the lower panels of Figures~\ref{d68fig} and~\ref{d68fig2}. Note that these parameters do not change much if we exclude the streak-pointed images, indicating that the two navigation methods are not biased relative to each other. In both data sets, we find a semi-major axis between 67625 and  67630 km, and an $ae$ $\sim$25 km.

\subsection{Evidence for variations in D68's mean radial position}
\label{drift}

\begin{figure}
\centerline{\resizebox{3.5in}{!}{\includegraphics{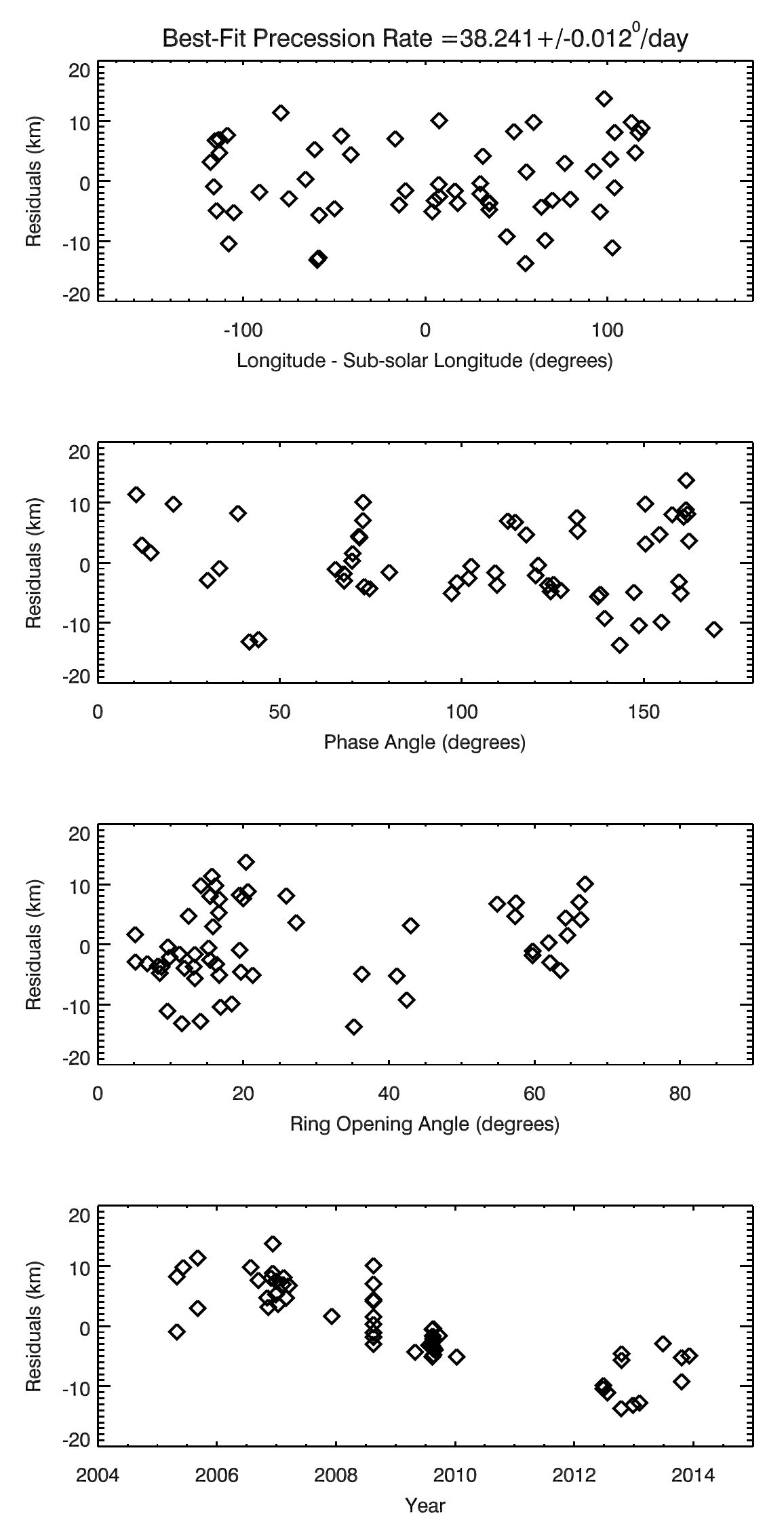}}}
\caption{The residuals from the best-fit eccentric model for all the data plotted  versus various parameters. There are no clear trends with longitude relative to the sun, phase angle or ring opening angle. However, a strong correlation with time is evident.}
\label{d68res}
\end{figure}

The residuals from the above best-fitting model are $\sim$7 km, which is much larger than the $rms$ dispersion of estimates within each image sequence. While high-resolution images once revealed a secondary peak $\sim$15 km from the main ringlet \citep{Hedman07}, such structures are unlikely to be responsible for the excess dispersion around these fits because such features would produce variations within individual movie sequences that are not observed (see Figure~\ref{d68poscomp} above). 

In principle, the excess dispersion around the best-fit model could reflect some systematic error in the navigation of images obtained at different viewing geometries. However, it could also reflect some slowly-evolving pattern not accounted for in our simple model. To examine these possibilities, we plot the residuals from the best-fit model as functions of the observed longitude (relative to the sub-solar longitude), ring-opening angle and phase angle in Figure~\ref{d68res}, and find no strong correlation with any of these parameters.  The lack of a trend with longitude relative to the Sun implies that this ringlet does not have a substantial component in its eccentricity that is forced by solar-radiation pressure, a phenomenon that has been observed in other dusty ringlets \citep{Hedman10, Hedman13}.  The uniform distribution of residuals with phase angle implies that particles with different sizes or light-scattering properties are not spatially segregated within the ring, which is consistent with the similar longitudinal brightness variations in Figure~\ref{d68intcomp}. Finally, the lack of any obvious increase in the $rms$ residuals at small opening angles indicates that this ringlet does not possess a detectable inclination (however, note that most of these observations were taken near the ring's  ansa, where vertical displacements should produce the smallest apparent radial displacements).

On the other hand, Figure~\ref{d68res} reveals  a clear trend in the residuals with observation time. Indeed, returning to Figures~\ref{d68fig} and~\ref{d68fig2}, we can see that the observations taken earlier in the Cassini mission give systematically larger radii than those taken later on. This trend was unexpected, and suggests that the effective semi-major axis of D68 has been drifting slowly inwards during the Cassini mission. Table~\ref{d68sep} shows fits to four sub-sets of the data, corresponding to four different time periods. These fits show no significant change in the ringlet's eccentricity between 2005 and 2013, but a steady reduction in the fitted semi-major axis. If we assume the semi-major axis follows a linear trend with time, then $da/dt =-2.38\pm0.39$ km/year. This decrease in semi-major axis should cause the ringlet's precession rate to increase by a factor of roughly 1$^\circ$/day/year. Such a change in the precession rate is consistent with the fitted pericenter locations advancing by about 15$^\circ$ between 2007 and 2013 (see Table~\ref{d68sep}).

\begin{figure}
\resizebox{6in}{!}{\includegraphics{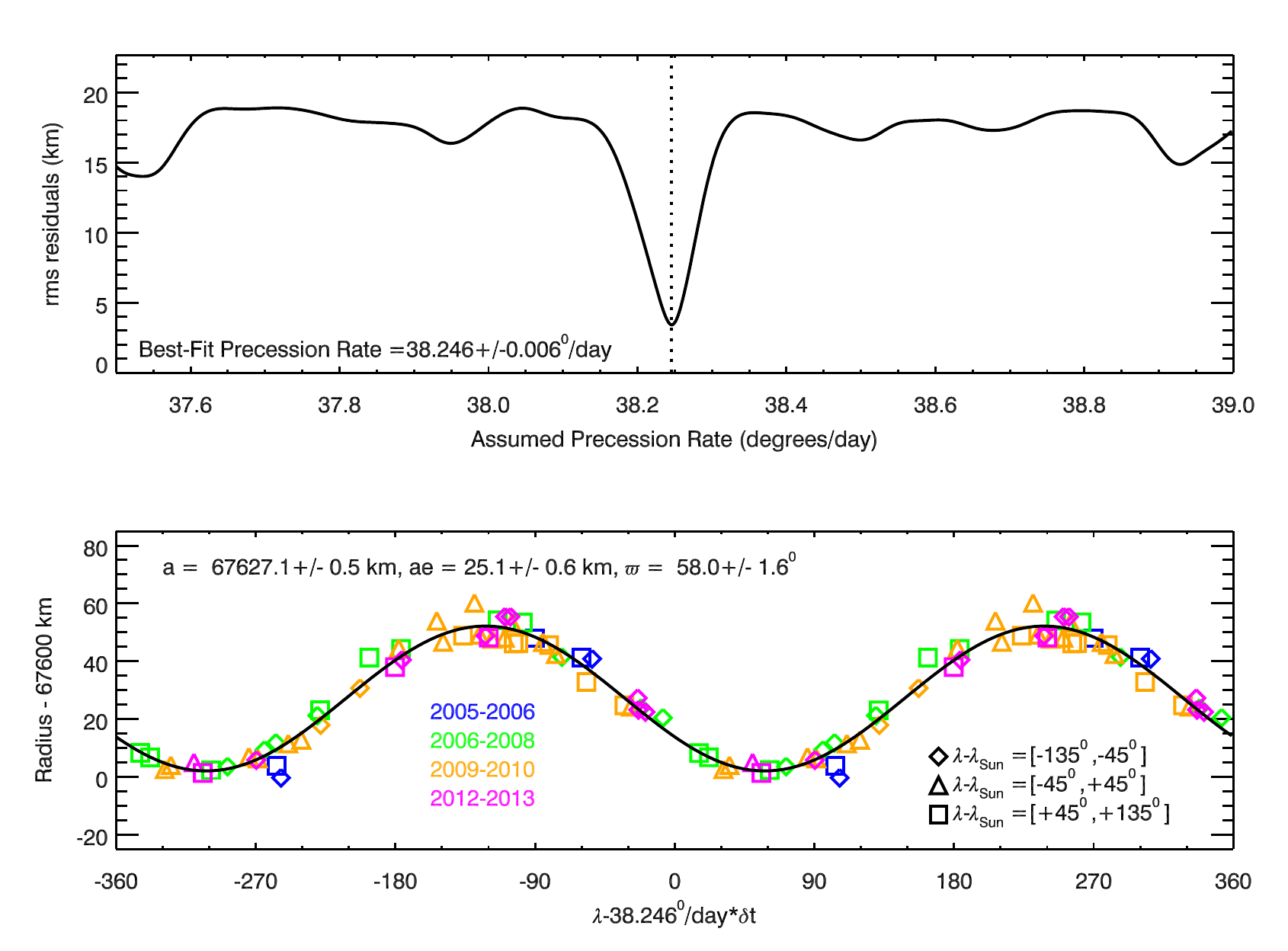}}
\caption{Fits of D68's radial position to a precessing-eccentric-ringlet model after removing a linear trend $da/dt =-2.38$ km/year. The top panel shows the $rms$ of the fit-residuals as a function of the assumed pattern speed. The best-fit pattern speed occurs at 38.246$^\circ$/day, where the $rms$ variations are roughly 3 km. The bottom panel plots the observed data and the best fitting model. Note the fitted semi-major axis $a$ is at an epoch time of 2009-185T17:18:54UTC or 300,000,000 TDB.  Different symbols indicate the observed longitude relative to the Sun and the colors designate the observation date. Note the much tighter dispersion in the residuals compared with Figure~\ref{d68fig}.}
\label{d68figx}
\end{figure}

\begin{figure}
\centerline{\resizebox{3.5in}{!}{\includegraphics{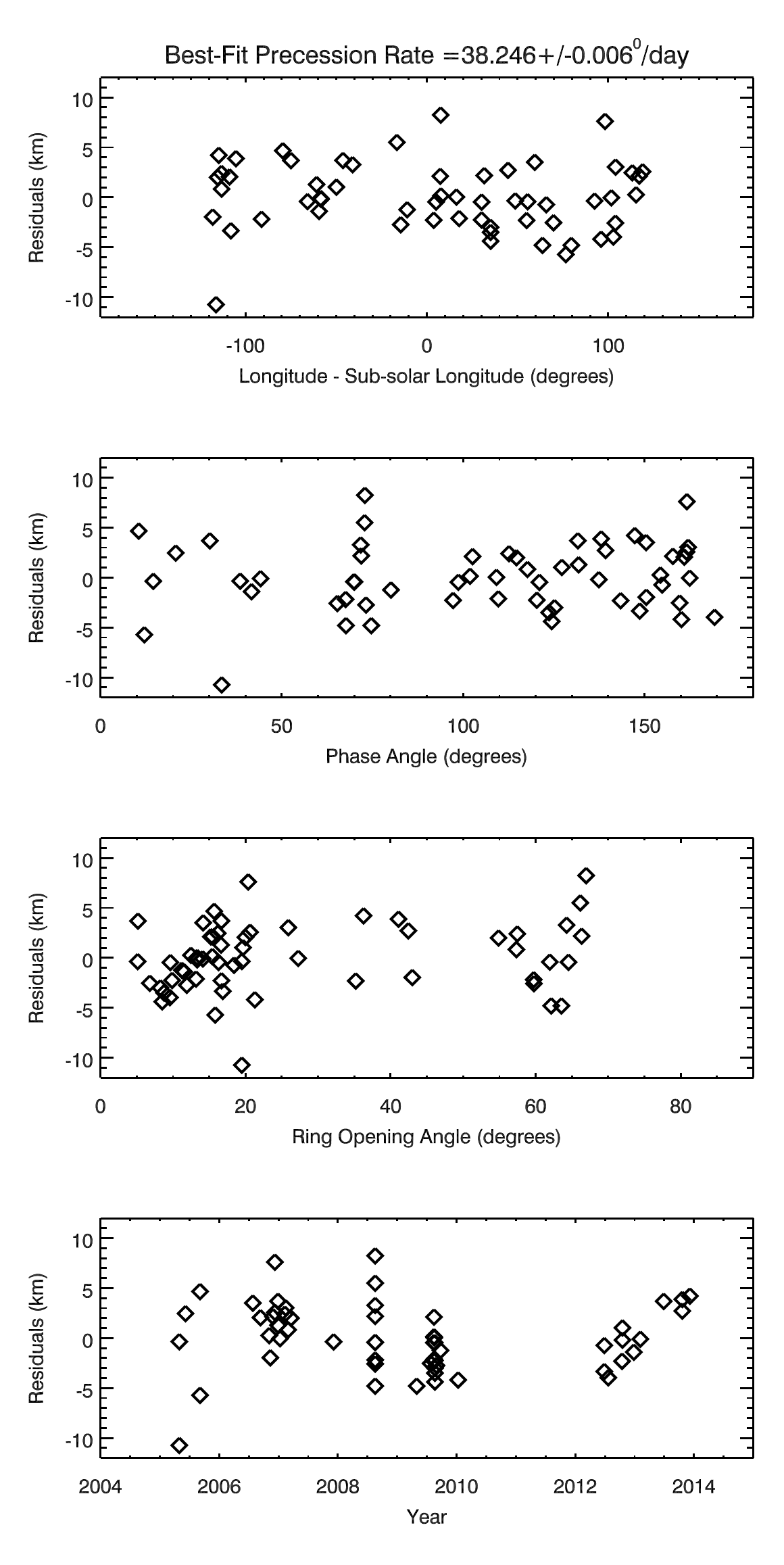}}}
\caption{Plots of the residuals from the best-fit eccentric model for all the data as functions of various parameters, after removing a linear trend $da/dt =-2.38$ km/year. No clear trends are evident with any of these parameters after removing this trend.}
\label{d68resx}
\end{figure}

If we subtract out this linear trend from the radius estimates by replacing $r$ with  $r+2.38\delta t$ in Equation~\ref{rmod}\footnote{For simplicity, these fits do not account for the predicted changes in precession rate induced by the semi-major axis changes. If we account for these phenomena, the pericenter longitude at epoch changes by 10$^\circ$, and the changes in the other parameters are negligible ($\dot\varpi$ changes by less than 0.002$^\circ$/day).}, then the $rms$ dispersion in the position estimates reduces to only 3 km around the best-fit model (see Figure~\ref{d68figx}), and there is still no observable trend with  longitude relative to the sun, phase angle or ring opening angle (see Figure~\ref{d68resx}). Thus only a trend with time can be clearly detected in these data.

\begin{table}
\caption{Fit parameters for different data subsets taken at different times (all assuming $\dot\varpi=38.241^\circ$/day)}
\label{d68sep}
\begin{tabular}{|c|c|c|c|c|}\hline
Time-Span & Mid Time & a & ae & $\varpi_o$ \\ \hline
2005.0-2006.0 & 2005.49 & 67633.3$\pm$3.2 km & 24.9$\pm$7.7 km & 70.2$^\circ\pm27.9^\circ$ \\
2006.0-2008.5 & 2007.02 & 67635.1$\pm$1.8 km & 26.4$\pm$2.6 km & 50.1$^\circ\pm5.3^\circ$ \\
2008.5-2011.0 & 2009.27 & 67626.6$\pm$1.5 km & 25.6$\pm$2.0 km & 53.0$^\circ\pm4.8^\circ$ \\
2012.0-2014.0 & 2013.09 & 67618.8$\pm$2.0 km & 24.7$\pm$2.6 km & 68.3$^\circ\pm7.3^\circ$\\
\hline
\end{tabular}
\end{table}

\begin{table}
\caption{D68 observations obtained by Voyager}
\label{d68voy}
\resizebox{6in}{!}{\begin{tabular}{|c|c|c|c|c|c|c|c|c|c|}\hline
Image & No. & UTC & Inertial & Phase & Emission & Sub-Solar & D68 & Radial & rms \\
 & & Time & Longitude & Angle & Angle &  Longitude & Radius & Resolution & Variations \\ \hline
3494650 &   1 & 1980-318T02:01:22 & 344.0$^\circ$ & 156.0$^\circ$ &  106.0$^\circ$ & 232.0$^\circ$ &  67574 km & 20.0 km & --- \\
4400753 &   1 & 1981-238T05:12:47 &  125.0$^\circ$ & 164.0$^\circ$ &  102.0$^\circ$ & 241.0$^\circ$ &  67609$^a$ km &  10.0 km & --- \\
\hline
\end{tabular}}
$^a$ Could shift up to 30 km inwards if D72 is assumed to be circular (see text).
\end{table}

While a linear decrease in $a$ with time improves the fit to the available Cassini data, earlier Voyager observations indicate that the long-term evolution of this ringlet's radial position is more complex. Voyager 1 and 2 each detected D68 in a single image \citep{Showalter96}. Table~\ref{d68voy} provides the pertinent geometric information for these two Voyager observations, along with position estimates for D68 derived from a revised analysis of the relevant images using the most recent SPICE kernels. The Voyager 1 image 3494650 contained 17 background stars that could be used for navigation, and the derived radial position of D68 in this image is 67,574 km, which is very close to previously published estimates \citep{Showalter96}. By contrast, the Voyager 2 image 4400753 only contained 3 star streaks, which makes the image navigation more uncertain. If we use the best-fit geometry to the star streaks, then we obtain a position estimate of 67,609 km, some 35 kilometers exterior to the Voyager 1 position.


We suspect that the revised position estimate for D68 from the Voyager 2 image is inaccurate for two reasons.
First, if D68 had a precession rate between 38.24$^\circ$/day and 38.25$^\circ$/day, then  the two Voyager images observed almost the exact same mean anomaly in the eccentric ringlet (i.e $|\delta \lambda - \dot{\varpi}\delta t|<5^\circ$).  The 35-km difference between the two Voyager position estimates is therefore somewhat surprising. While it is possible that D68 had a different shape in the 1980s, the narrow ringlet D72 was also found further from the planet in the Voyager 2 image than it was in the Voyager 1 image  (71,731 km compared to 71,704 km). Both ringlets being displaced by roughly the same amount in the same direction would be quite a coincidence, and strongly hints that  some systematic error is affecting the geometry of the Voyager 2 image, and it is likely that D68 was close to 67,580 km in both Voyager images. 

If we assume the 38.241$^\circ$/day precession rate that best fits the full Cassini data, and extrapolate back to the Voyager epoch, then both the Voyager images viewed regions within a few degrees of D68's pericenter (If we instead use the 38.246$^\circ$/day rate that best fits the de-trended data, we find these data occur roughly 30$^\circ$ from pericenter). If we also suppose that D68's eccentricity remains constant and extrapolate the linear semi-major axis drift back over 30 years, then we would expect D68 to have a pericenter radius between 67,640 and 67,680 km when the Voyager images were taken. This is more than 50 km exterior to the more reliable Voyager 1 position estimate (and even outside the questionable Voyager 2 measurement). Hence either the ringlet's  shape changed dramatically between the Voyager and Cassini epochs, or D68's semi-major axis did not drift inwards at a purely constant rate over the last 35 years.  Given that the variations in $a$ are much larger than those in $ae$ in the Cassini data, we regard the latter option as more likely.


A closer inspection of the residuals in the bottom panel of Figure~\ref{d68res} provides further evidence that the ringlet's radial motion is more complex than a constant radial drift. The data from 2005 fall below an extrapolation of the trend seen in the data obtained between 2006 and early 2013, while the data from late 2013 fall above this same trend. This suggests that the ringlet's  semi-major axis may have reached its maximum value in 2006 and its minimum value in 2013, which would imply an oscillation period of  14-15 Earth years, or about half a Saturn year. Further observations by Cassini should reveal whether this apparent oscillation is real, and provide proper constraints on its amplitude and period. We should also note that the Cassini observations cover a time period exactly one Saturn year after the Voyager encounters, and the Voyager 1 radial position measurement is beyond the range of any Cassini observations to date. Thus D68's motions might also include trends that extend over timescales longer than Saturn's seasonal cycle.

\section{Discussion}
\label{discussion}

The above analysis reveals that (1) D68 possesses substantial longitudinal brightness variations that seem to drift around the planet at about 1750$^\circ$/day, (2) D68 has a large eccentricity and its orbit precesses around the planet at about 38.24$^\circ$/day, and (3) D68's mean radial position shifts in and out on decadal timescales. 
These phenomena contain information about the ringlet's dynamical environment, including higher-order components of Saturn's gravitational field, but we have not yet been able to uniquely identify the physical processes responsible for sculpting and perturbing this ringlet. Hence the following discussion is only a preliminary investigation of D68's potential dynamical implications. 

We begin by considering the origin of D68's visible dust in Section~\ref{origin}, and argue it is unlikely that D68 is simply tracing out the orbit of a single larger object, which implies that the visible material in D68 is being confined in radius somehow. Section~\ref{pertaz} discusses how radius-dependent and time-variable azimuthal forces might be able to explain this radial confinement, along with the ringlet's finite eccentricity and radial migration. Section~\ref{pertid} discusses some physical mechanisms that could produce such forces. Finally, Section~\ref{prec} discusses D68's precession rate and how it could be used, in conjunction with data from other  close-in rings, to further constrain Saturn's gravitational field.

\subsection{The source of D68}
\label{origin}

D68 is a particularly puzzling feature because it is a narrow ringlet that is not near an obvious dust source. Outside of the D ring, dusty ringlets less than 100 km wide are only found within nearly-empty gaps in the main rings (e.g., the Encke Gap or Laplace Gap, see Hedman {\it et al.} 2010, 2013) or close to the orbit of one of Saturn's smaller moons (e.g., Aegaeon, Anthe and Methone, see Hedman {\it et al.} 2009). All these other ringlets therefore occur in the vicinity of larger objects. This is sensible because micron-sized particles in the Saturn system can be eroded by sputtering or micrometeoroid impacts on time scales well less than 1000 years \citep{BHS01}, so the visible dust in these ringlets needs to be constantly re-supplied from larger source bodies. 


Perhaps the simplest way to explain D68 would be  to posit a single moonlet like Anthe or Methone embedded within the ringlet. In this case, the ringlet could simply trace the moon's orbit, which could have a finite eccentricity and may exhibit some long-term semi-major axis librations. Furthermore, the persistent longitudinal brightness variations could be attributed to horseshoe or tadpole-like particle motions around the moon's Lagrange points. Unfortunately, examinations of low-phase movie sequences have thus far failed to reveal a moonlet in the vicinity of D68. Furthermore, the existence of such a moon would be difficult to reconcile with D68's location. D68 is well within the Roche limit of any ice-rich object, so an isolated solid object should not be able to form via gravitational aggregation near this ringlet. Even if a suitable object somehow drifted into this region from farther out, meteoroid impacts would knock off material, leading to a population of ever-smaller objects. Thus, we strongly suspect that D68 is not generated by a single moonlet.

If D68 is not generated by a single moonlet, then there are two plausible alternative sources for the visible material: either (1) the dust is produced locally by a population of multiple source bodies or (2) the dust is produced elsewhere and becomes trapped in this region. In both these scenarios, there must be some active process that confines material  around D68. For the first option, the relevant source bodies themselves would need to be confined to a narrow annulus, and so the confinement process would influence both large and small particles. For the second option, we note that D68 is embedded in a broad sheet of dust extending interior to the C-ring's inner edge, which likely consists of particles spiraling into the planet under the influence of plasma drag (see below). If D68 represents a concentration of dust-sized particles derived from elsewhere in the ring, then the relevant confinement mechanism only needs to influence the dynamics of these tiny grains.

At present, we have no direct evidence that there are  larger source bodies embedded in D68. For example, none of the images show any discrete bodies within the ringlet. However, the three highest-resolution, lowest-phase images of D68 obtained back in 2005 (and described by Hedman {\it et al.} 2007) may provide some indirect evidence for local source bodies.  Two of these images  revealed sub-structure in D68's radial profile, and in one D68 contained two brightness peaks separated by 15 km. This radial structure could potentially reflect gravitational or collisional interactions between the D68 ringlet particles and some larger source body embedded in the ring. Indeed, multi-stranded patterns in Saturn's F ring can be attributed to both known moons like Prometheus and smaller embedded objects \citep{Charnoz05, Murray05, Murray08, Charnoz09, Attree12}. Also, the arc of debris surrounding Anthe sometimes appears to be sculpted into multiple strands in the vicinity of that moon \citep{Hedman09}.  

Unfortunately, secondary strands  are not clearly visible in any images of D68 obtained after 2005. Some of these later observations have comparable resolution to the 2005 measurements, but were obtained at higher phase angles and/or lower signal-to-noise, so the absence of secondary features in these images could just  be an artifact of the viewing conditions. Hence the distribution and motion of these small-scale radial features are still unclear. Furthermore, it is difficult to extract any information about any putative embedded objects' properties based on the sparse available data. If these secondary strands were produced by gravitational perturbations from an embedded object, then the observed radial separations between these strands would be controlled by the object's Hill sphere (i.e., its mass). However, embedded objects can also produce secondary strands via collisions with pre-existing ring particles. The morphology and locations of such impact-generated strands depend mainly on the orbits of the ringlet particles and object, and the separations between these strands and the main ringlet can be much larger than the moon's Hill sphere \citep{Charnoz05, Murray05, Murray08, Charnoz09}. The limited high-resolution data on the rings' morphology therefore do not yet provide firm limits on the size or number of source bodies in D68. 


\subsection{Perturbing forces acting on D68}
\label{pertaz}

Even without evidence whether or not D68 contains larger particles, the slow radial motion of D68 provides clear evidence that some process is causing the orbital elements of its constituent particles to change over time. Indeed, the observable trends in the ringlet's semi-major axis and eccentricity furnish some clues regarding the magnitude and direction of these forces.

If we make the reasonable assumption that the ringlet's shape represents the average orbital elements of its constituent particles, then we may infer that over the course of the Cassini mission these particles had a typical semi-major axis drift rate $da/dt=-2.38\pm0.39$ km/year and an eccentricity change that is consistent with zero ($ade/dt=-0.23\pm0.57$ km/year). Standard perturbation theory provides expressions for the evolution of a particle's orbital semi-major axis and eccentricity in terms of a generic perturbing force \citep{Burns76}. For particles on nearly circular orbits ($e<<1$), the relevant equations can be approximated as follows:
\begin{equation}
\frac{da}{dt}=2an\left[\frac{F_r}{F_G}e\sin f+\frac{F_\lambda}{F_G}(1+e\cos f) \right],
\label{apert}
\end{equation} 
\begin{equation}
\frac{de}{dt}=n\left[ \frac{F_r}{F_G}\sin f +2\frac{F_\lambda}{F_G}\cos f\right],
\label{epert}
\end{equation}
where $n$ and $f=\lambda-\varpi$ are the particles' orbital mean motion and true anomaly, respectively, $F_r$ and $F_\lambda$ are the radial and azimuthal components of the perturbing force, and $F_G=GM_Sm/a^2$ is the central gravitational force from the planet ($G$ being the universal gravitational constant, while $M_S$ and $m$ being the masses of Saturn and the particle, respectively). Note that for the ringlet's semi-major axis to drift in one direction over several years, the average of $da/dt$ over all $f$ is nonzero, which can occur if either one of the following two conditions are satisfied.
\begin{itemize}
\item The azimuthal component of the perturbing force has a non-zero orbit average.
\item The perturbing force varies over the course of an orbit such that $F_r\cos f$ and/or $F_\lambda \cos f$ have non-zero mean values.
\end{itemize}
In fact, the first of these options is the most likely one. Note that these two equations contain some of the same terms, and so we can in fact combine these two formulae to yield the following expression for $da/dt$ in terms of $ade/dt$ and $F_\lambda$:
\begin{equation}
\frac{da}{dt}=2ae\frac{de}{dt}+2an\frac{F_\lambda}{F_G}(1-e\cos f).
\end{equation}
For D68, $e=0.00037$,  $da/dt=-2.38\pm0.39$ km/year and $ade/dt=-0.23\pm0.57$ km/year, hence $da/dt >> ae(de/dt)$ and the second term on the right-hand side of the above expression must be the dominant one. Furthermore, since $e<<1$, a constant $F_\lambda$ provides the most efficient way to cause long-term changes in the semi-major axis. If we assume $F_\lambda$ is strictly constant, then we may solve the above equation for $F_\lambda/F_G$ and determine that between  2005 and 2012 the particles in D68 felt an average azimuthal force $F_\lambda \simeq -1.7\times10^{-9}F_G$. 

Of course, this force cannot be strictly constant or else the ringlet would have moved steadily inwards between the Voyager epoch and today. Instead, it appears that the ringlet moved outwards between the Voyager and Cassini epochs, and may be beginning to do so again in 2013. Hence during some time periods the perturbing force must have acted to accelerate the ring particles in their direction of motion (i.e., $F_\lambda$ was positive). Since particles at all longitudes moved back and forth through the same region, this implies that the strength and magnitude of this perturbing force changes with time. 

We  may also interpret the ringlet's narrow radial width and eccentricity as the response of the ringlet particles to azimuthal forces. One way to keep the ringlet narrow is to have a perturbing force that drives the semi-major axis of any given ring particle towards a particular value. Again, Equation~\ref{apert} indicates that the constant azimuthal force option will be most efficient at generating semi-major axis drifts (that term does not include a factor of $e$, which is small for D68). Thus the ringlet could be confined by a force that accelerates the particles along their direction of motion when they get closer to the planet, and slows them down if they stray too far outside D68's mean radius. That is, $F_\lambda \propto (r_0-r)$, where $r_0$ is close to D68's (slowly time-variable) semi-major axis. 

Intriguingly, an azimuthal force of this type would also help maintain the ringlet's finite eccentricity. If we assume the particle's semi-major axis $a\simeq r_0$, then $F_\lambda \propto ae\cos f$. Inserting this force into Equation~\ref{epert} and averaging over a single orbit, we find $de/dt \propto +e$. This makes sense, since this force will accelerate the particle near its orbital pericenter and slow it down near its orbital apocenter, which will cause the orbit's eccentricity to grow. 


Based on the above considerations, it is tempting to think that the ringlet's radial migration, narrow width and finite eccentricity can all be ascribed to a radius- and time-dependent azimuthal force. However, one should  recall that other processes could be involved. In particular, we should not neglect the potential role of inter-particle interactions. While D68's normal optical depth is low ($<<10^{-3}$ , consistent with the non-detection of D68 in occultation data), and so direct collisions between ring particles should be rare, this does not necessarily make them negligible. For example, inter-particle interactions could explain the lack of obvious correlations between the ringlet's measured position and the observed phase angle. If each particle in D68 moved completely independently of all the others, then this observation would imply that the forces sculpting the ringlet do not segregate particles by size, and thus must be purely gravitational. However, tenuous dusty ringlets in the Encke Gap and the Cassini Division exhibit coherent motions due to non-gravitational forces like solar radiation pressure \citep{Hedman10, Hedman13}. While the mechanism that organizes  the particle motions in these ringlets remains uncertain, it likely involves inter-particle interactions. These interactions could potentially even excite free eccentricities and prevent the ringlet's dispersal \citep{Hedman10, Lewis11}. We therefore regard the radial migration as the clearest evidence for an outside force perturbing the ring, and furthermore the observed drift rate provides a rough estimate of that force's magnitude.

\subsection{Candidate perturbing forces}
\label{pertid}

Assuming that the typical ring particle in D68 feels an average azimuthal force amounting to $\sim2\times10^{-9} F_G$, we may ask what could generate that perturbing force. The options we will consider below include both gravitational perturbations associated with co-rotation resonances and non-gravitational forces like plasma drag. While some of these forces have the required strength to produce drifts similar to those observed, none of them on their own seems to be able to explain all of D68's radial motions.

\subsubsection{Gravitational perturbations from co-rotation resonances}

Gravitational forces can induce strong azimuthal accelerations in the vicinity
of co-rotation resonances, similar to those responsible for producing the arcs in the Anthe, Methone and G rings. A first-order co-rotation eccentricity resonance produces longitudinal forces with $F_\lambda/F_G$ of order the moon's eccentricity times the ratio of the moon's mass to Saturn's mass.  Thus resonances with Saturn's larger satellites would have sufficient strength to produce the radial motions observed. Unfortunately, no such resonances occur anywhere near D68, so it seems unlikely that gravitational pulls from Saturn's various moons are responsible for maintaining D68.\footnote{The only satellite resonance near D68 is the 3:1 Inner Lindblad Resonance with Atlas, which occurs at 67,564 km if we use the gravitational field parameters given by \citet{Jacobson06}. However, this resonance is unlikely to be responsible for any of the structures in D68. A 3:1 resonance produces an $m=2$ structure that rotates around the planet at roughly one half the local orbital rate. This resonance therefore cannot be responsible for the observed eccentricity. Also, being a Lindblad resonance, it is not likely to produce longitudinal brightness variations or slow semi-major axis shifts.} 

However, Saturn's moons are not the only possible source of gravitational perturbations;  asymmetries or oscillations inside the planet can also influence ring-particle's orbital properties \citep{Marley91, MarleyPorco93}. Recently, \citet{HN13} determined that certain density waves in Saturn's C ring have the right symmetry properties and pattern speeds to be generated by resonances with low-order normal-mode oscillations inside Saturn.  Furthermore, detailed studies of these waves yielded precise measurements of the pattern speeds of six different normal modes, which all fell between 1650$^\circ$/day and 1900$^\circ$/day, which is comparable to the ring-particles' expected mean motion near D68. Thus perhaps D68 could be sculpted and maintained by a form of co-rotation resonance with Saturn.

The simplest planetary co-rotation resonances occur where the mean motion of the ring particles' $n$ equals the pattern speed $\Omega_P$ of the (prograde) normal mode. At these locations, the torques from an $m$-lobed distortion in the planet give rise to $m$ stable locations where material could be trapped. Unfortunately, none of the normal modes identified to date have a pattern speed that exactly matched the expected mean motion of the D68 particles. The closest options are an $m=2$ pattern with $\Omega_P=1779.5^\circ$/day and an $m=3$ pattern with $\Omega_P=1736.6^\circ$/day, which are over 14$^\circ$/day from the expected mean motion of D68, and also do not correspond to any of the possible values for the mean motion of the longitudinal brightness variations observed in D68 (see Figure~\ref{d68mark}). 

While it remains possible that an as-yet unidentified planetary oscillation could have the right pattern speed to produce a resonance in the vicinity of D68, there is a more generic problem with invoking co-rotation resonances to explain D68's radial motion. Any such resonance produces forces that depend not only on radius, but also co-rotating longitude. These resonances therefore tend to produce longitudinally confined arcs, and not complete rings like D68. Indeed it is not clear how a co-rotation resonance could cause an entire ring to move in and out coherently.

\subsubsection{Non-gravitational perturbations due to plasma drag}

One potential non-gravitational perturbing force that could be acting on D68 is plasma drag, which arises due to momentum exchange between the ring particles and the plasma ions. These ions are strongly coupled to Saturn's magnetic field and thus move around the planet once each planetary rotation period, which is around 10.5 hours. By contrast, the orbit period of D68 particles is around 5 hours, which means the plasma ions move around the planet more slowly than the ringlet particles. Hence momentum exchange with the ions causes the ringlet particles to lose energy and spiral in towards the planet. For a suitably low-density plasma, the magnitude of the drag force is given by the simple expression \citep{Grun84}:
\begin{equation}
|F_{drag}|=\pi s^2n_im_iw^2,
\end{equation}
where $s$ is the particle size (radius), $n_i$ is the ion number density, $m_i$ is the ion mass and $w$ is the ion's velocity relative to the particles. Thus the force ratio can be expressed as:
\begin{equation}
|F_{drag}/F_G|=\frac{3n_im_ia}{4\rho_g s}\left(1-\frac{\Omega_S^2}{n^2}\right),
\end{equation}
where $\rho_g$ is the mass density of the ringlet particles and $\Omega_S=2\pi/10.5$ hours =  0.00017/s is the planet's rotation frequency. For D68's $a$ and $n$, this expression can be written as:
\begin{equation}
|F_{drag}/F_G|\simeq6\times10^{-11}\left(\frac{n_i}{cm^{-3}}\right)\left(\frac{m_i}{amu}\right)\left(\frac{1 g/cm^{3}}{\rho_g}\right)\left(\frac{1 \mu m}{s}\right).
\end{equation}
Radio occultations of Saturn's ionosphere indicate that the electron density near Saturn's equatorial plane around D68's position is between 10 and 100 electrons/cm$^3$ \citep{Nagy06, Kliore09}. Assuming hydrogen ions have a comparable number density in this region, and that the typical particle size in this region is of order a few microns, then plasma drag could be sufficient to produce the radial migration observed in the Cassini data.

However, plasma drag cannot explain how the ringlet moved outwards between the Voyager and Cassini epochs. Any plasma co-rotating with Saturn's magnetosphere will orbit the planet more slowly than the dust grains, so any momentum exchange with the plasma must cause the dust to spiral inwards. Worse, the ion density should increase rapidly as the particles approach the planet, which means particles near the inner  edge of the ringlet should move inwards faster than the particles near its outer edge. This will tend to disperse the ringlet, rather than confine it. This implies that some other force is acting on the grains besides plasma drag. This force must be accelerating dust grains along their direction of motion and must be able to balance or overwhelm plasma drag in these regions. 

One possible way to balance plasma drag would be to have the visible D68 particles exchange momentum with material moving around the planet faster than the local Keplerian rate. Sub-micron particles with substantial positive charges could do this because Saturn's magnetic field would produce an additional central force for these particles, leading to a faster orbital mean motion. However, not only would this require an unseen population of tiny particles, it is also unclear if the plasma conditions in the D ring would positively charge those grains.  

Finally, we may note that if the semi-major axis does in fact oscillate with a period of around 14 or 15 years, this might provide another clue regarding the forces driving this motion. For example, seasonal processes are likely involved if the period turns out to be exactly half a Saturn year. Similarly semi-annual oscillations have been observed in Saturn's upper atmosphere \citep{Orton08}, which demonstrate that some aspects of D68's environment could be changing on this timescale. However, at present it is unclear whether there is any sensible physical connection between D68 and any relevant atmospheric or seasonal phenomenon.

\subsection{D68's Precession Rate and Constraints on Saturn's Gravity Field}
\label{prec}

The Cassini observations yield a very precise measure of D68's apsidal precession rate $\dot{\varpi}=38.243\pm0.008^\circ$/day (the error here includes systematic uncertainties associated with the semi-major axis evolution). This rate not only provides further information about the ringlet's dynamical environment, it is also a unique probe of Saturn's internal structure. In the small-eccentricity limit, a particle's precession rate is given by the perturbation equation:
\begin{equation}
\frac{d\varpi}{dt}=n\left[ -\frac{F_r}{F_G}\cos f +2\frac{F_\lambda}{F_G}\sin f\right].
\end{equation}
Note that this expression will average to zero for an azimuthal force $F_\lambda$ that is either constant or proportional to $r_0-r$, so the perturbations that would be most efficient at confining the ringlet, causing it to migrate, or exciting its eccentricity do not influence this parameter. Instead, the precession rate is most sensitive to radius-dependent {\em radial} forces, One well-known source of such forces are the higher-order components of Saturn's gravitational field. Indeed, previous studies of eccentric structures in the inner C ring have already demonstrated that ringlet precession rates  can place important constraints on Saturn's gravitational field \citep{NP88}. A simultaneous fit to the precession rates of ringlets in both the D and C rings therefore promises to be very informative. While such a comparative investigation is beyond the scope of this paper, we will present a preliminary study to illustrate some of the challenges involved in extracting information about the planet's gravity field from these data. 

One obvious issue with using D68 to constrain the planet's gravity field is that we do not yet know if the processes responsible for the ringlet's eccentricity or radial migration are also subtly perturbing its precession rate. In principle, this concern can be addressed by comparing data from multiple eccentric ringlets and determining if they can all be fit with a single coherent model of the planet's gravity field. Hence we can leave these particular extensions for a later work. 

Another issue is that D68's observable precession rate does not place a straightforward constraint on any of the parameters in the standard expansion of Saturn's gravitational field. Normally, the axisymmetric component of Saturn's gravitational potential versus radius $r$ is written in terms of the series:
\begin{equation}
V=-\frac{GM_s}{r}\left[1-\sum_{i=2}^\infty J_i(R_p/r)^iP_i(\sin\theta)\right],
\end{equation}
where $\theta$ is latitude, $R_p=$ 60,300 km is the planetary radius, $P_i$ are Legendre polynomials of degree $i$ and $J_i$ are numerical coefficients. Note that for a fluid planet only even $i$ should have non-negligible coefficients, so typically the planet's higher-order gravity can be described by the coefficients  $J_2, J_4, J_6...$. The apsidal precession due to these factors can be most easily written down if we recall that  $\dot{\varpi}=n-\kappa$, where $n$ and $\kappa$ are the orbital mean motion and radial epicyclic rates, respectively. These two rates can be expressed as the following functions of the particles' semi-major axis $a$ \citep{MD99}:
\begin{equation}
n^2=\frac{GM_s}{a^3}\left[1-\sum_{i=2}^\infty (2i+1)J_i(R_p/a)^iP_i(\sin\theta)\right]
\end{equation}
\begin{equation}
\kappa^2=\frac{GM_s}{a^3}\left[1+\sum_{i=2}^\infty (2i+1)(2i-1)J_i(R_p/a)^iP_i(\sin\theta)\right].
\end{equation}
This expansion of the potential is perfectly functional when dealing with Saturn's moons and outer rings because the factor $(R_p/a)^i$ means that the lowest-order terms like $J_2$ and $J_4$ have the biggest effect on the observed precession rates. For features in the inner C ring or D ring, however, this expansion becomes problematic because $R_p/a$ is not much less than one and so many different terms can contribute to the precession rate. This problem was already identified in the context of the Titan ringlet by \citet{NP88}, but for D68 this problem becomes particularly  acute because it is so close to the planet $(R_p/a \simeq 0.89)$.

\begin{figure}
\resizebox{6in}{!}{\includegraphics{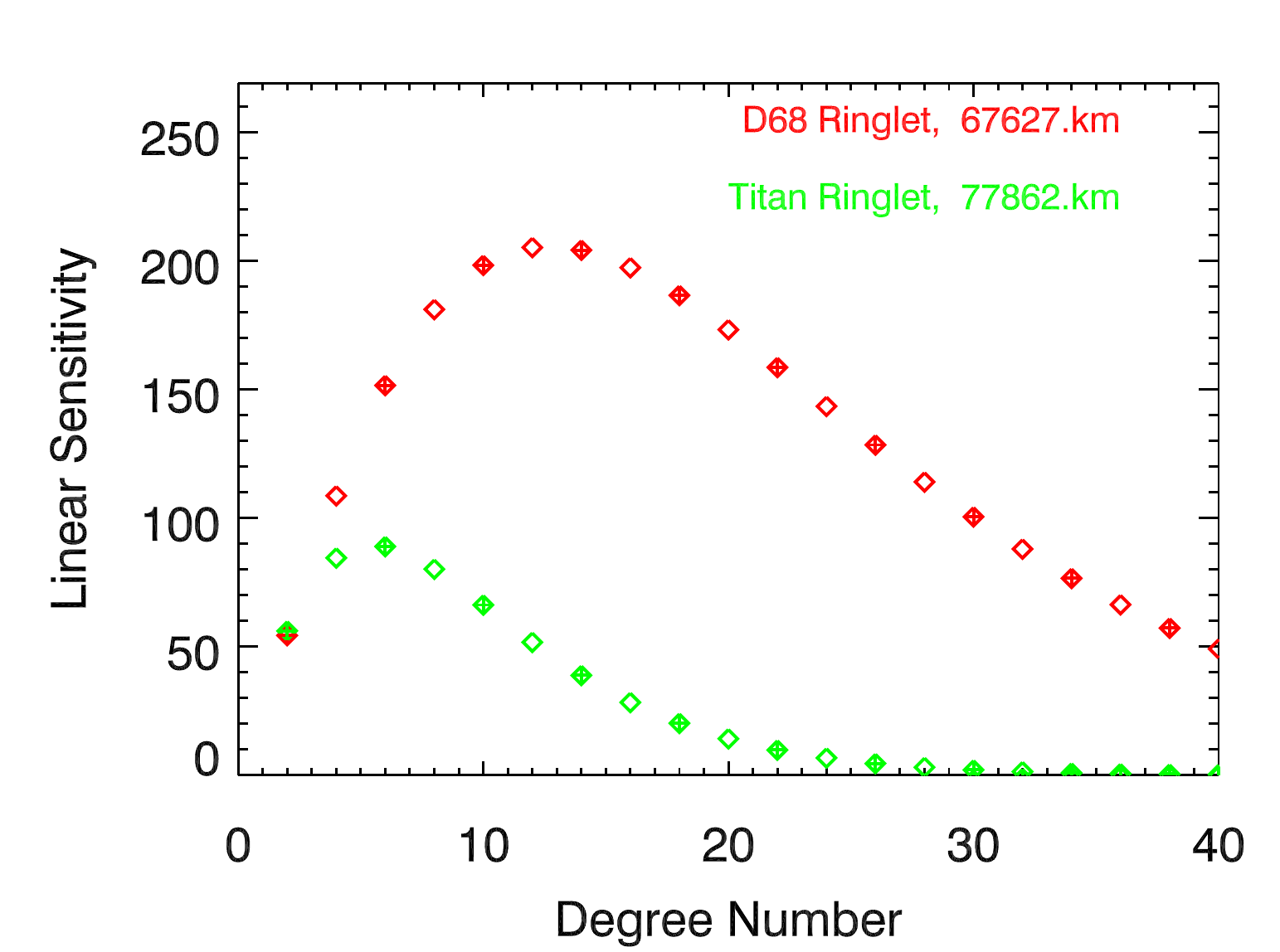}}
\caption{Linear sensitivity of D68's and the Titan ringlet's precession rates to various gravitational harmonics. Linear sensitivity is defined as the fractional change in the precession rate per small change in the gravitational harmonic from a nominal model. Solid and empty symbols correspond to positive and negative sensitivities, respectively. Note the large range of harmonics that can influence D68's precession rate.}
\label{sensplot}
\end{figure}

Figure~\ref{sensplot} shows the linear (fractional) sensitivity of D68's precession rate to a small change in any of the gravitational moments $J_i$: 
\begin{equation}
\mathcal{S}_i=\frac{1}{\dot{\varpi}}\frac{\partial \dot{\varpi}}{\partial J_i}.
\label{senseq}
\end{equation}
\citet{NP88} computed these coefficients for the Titan ringlet in the inner C ring, and 
found that while the ringlet's precession rate was most sensitive to $J_6$, it was actually sensitive to a broad range of coefficients between $J_2$ and $J_{20}$. D68 is even more extreme, being most sensitive to $J_{12}$, but having a broad peak extending well out beyond $J_{40}$.  These graphs clearly demonstrate that D68's precession rate alone cannot constrain any particular harmonic coefficient. Worse, even if we assume the low-order moments $J_2, J_4$ and $J_6$ are known from other observations, there are still a large number of combinations of higher-order moments that could reproduce any given precession rate. 

\citet{NP88} presented their constraint on Saturn's gravity field in terms of a linear combination of harmonic coefficients. A similar procedure can be applied to the D68 data here. In general, the constraint can be expressed using the following expression:
\begin{equation}
\frac{\dot\varpi_{obs}-\dot\varpi_{mod}}{\dot\varpi_{mod}}=\sum(J_i-J_{i,mod})\mathcal{S}_i,
\end{equation}
where $\mathcal{S}_i$ are the coefficients from Equation~\ref{senseq} above, $\dot\varpi_{obs}$ is the observed precession rate, and $\dot\varpi_{mod}$ is the computed precession rate at the appropriate semi-major axis assuming the model values for the harmonic coefficients $J_{i,mod}$. Assuming $a=67,628 km$, and using the \citet{Jacobson06} model for Saturn's gravity field (i.e., $J_{2,mod}=16290.71\times10^{-6}, J_{4,mod}=-935.85\times10^{-6}, J_{6,mod}=86.14\times10^{-6}, J_{8,mod}=-10\times10^{-6}$ and all other $J_{i,mod}=0$), the left-hand side of this equation becomes:
\begin{equation}
-0.029\pm0.002\pm0.006=\sum(J_i-J_{i,mod})\mathcal{S}_i.
\label{solution}
\end{equation}
Note the first error is the actual error in the measured precession rate, while the second represents a conservative $\pm10$ km error in the ringlet's effective semi-major axis
(i.e., $(\partial\dot\varpi_{mod}/\partial a)\delta a/\dot\varpi_{mod}$). The relevant $\mathcal{S}_i$ and $J_{i,mod}$ parameters are given in Table~\ref{params}. However, it should also be clear that this formalism does not provide much direct insight into the structure of the planet's gravitational field.

\begin{table}
\caption{The parameters used in Equation~\ref{solution}}
\label{params}
\centerline{\begin{tabular}{|c|c|c||c|c|c|} \hline
i & $\mathcal{S}_i$ & $J_{i,mod}$ & i & $\mathcal{S}_i$ & $J_{i,mod}$ \\ \hline
2 & +54 & 16290.71$\times10^{-6}$ & 22 & +158 & 0 \\
4 & -108 &  -935.85$\times10^{-6}$  & 24 & -143 & 0 \\
6 & +151 &  86.14$\times10^{-6}$ &26 & +128 & 0 \\
8 & -181 & -10$\times10^{-6}$ & 28 & -113 & 0 \\
10 & +198 & 0 & 30 & +100 & 0 \\
12 & -205 & 0& 32 &  -88 & 0 \\
14 & +204 & 0 & 34 &  +77 & 0 \\
16 & -197 & 0 & 36 & -66 & 0 \\
18 & +186 & 0 & 38 &+57 & 0 \\
20 & -173 & 0 & 40 & -49 & 0 \\
 \hline
\end{tabular}}
\end{table}

In order to gain a bit more physical insight into what D68 can tell us about Saturn's gravitational field, consider the following: For a giant planet like Saturn, two different phenomena can generate or modify the relevant gravitational moments (assuming a particular model for the planet's internal structure): (1) the planet's oblateness due to its rotation and (2) the dynamics of its zonal winds \citep{Kaspi10, Kaspi13a, Kaspi13b}. Hence we can try to create a toy model that roughly represents both of these phenomena.

For the terms generated by the planet's rotation, we use harmonic terms similar to those for a homogeneous Maclaurin spheroid. This object has gravitational moments $J_i$ given by the expression \citep{Hubbard12}
\begin{equation}
J_i=\frac{3(-1)^{1-i/2}}{(i+1)(i+3)}\left(\frac{\ell^2}{1+\ell^2}\right)^{i/2},
\label{macj}
\end{equation}
where $\ell$ is a measure of the oblateness of the planet. In this scenario, all the gravitational harmonics are determined by the single parameter $\ell$, greatly reducing the parameter space. Of course, Saturn is not homogeneous, so the above expression cannot give a reasonable approximation of even the lowest-order harmonics like $J_2$, $J_4$ and $J_6$. However, the above expression gives the appropriate first-order trend in the gravitational harmonics for a rigidly rotating planet, so we will use this expression as an approximation for the harmonic moments that are not yet well constrained by observations. For Saturn, $J_2$ and $J_4$ are well-measured, while $J_6$ is still somewhat uncertain \citep{Jacobson06}. We can therefore set $J_2$ and $J_4$ to their observed values, and have the remaining $J_i$ given by the following expression:
\begin{equation}
J_i=J_4\frac{35(-1)^{1-i/2}}{(i+1)(i+3)}\left(\frac{\ell^2}{1+\ell^2}\right)^{i/2-2}.
\label{macj}
\end{equation}
Hence this part of the planet's gravity field depends upon the parameters $J_2$, $J_4$ and $\ell$. In practice, instead of presenting precession rates as functions of $\ell$, we can express them in terms of $J_6=-5/9J_4\ell^2/(1+\ell^2)$, which is easier to compare to existing estimates of Saturn's gravity field.

For any reasonable value of $J_6$, the above series will converge towards zero rapidly, ($J_6/J_4 \sim 0.1$, so in general $J_{i+2}/J_i$ will be of comparable size). However, for real giant planets, the very high-order harmonics can be enhanced by orders of magnitude due to contributions from the planet's zonal winds \citep{Kaspi10}. In principle, the contribution of Saturn's zonal winds can be computed assuming a density profile and a scale height for the winds \citep{Kaspi10, Kaspi13a, Kaspi13b}. However, in the interest of simplicity we will instead model the winds' contribution to the planet's gravitational field using a single massive wire wrapped around Saturn's equator. While this is obviously a gross oversimplification, it can perhaps be justified because the dominant feature in Saturn's wind profile is its equatorial jet, and a massive wire on the equator can serve as a zeroth-order approximation of the mass anomaly associated with such a feature. Furthermore,  the wire model has the great advantage that the precession rate due to such a wire has been computed  \citep{Null81}.
\begin{equation}
\dot{\varpi}_w=\frac{n}{4}\frac{m_w}{M_s}(R_W/a) b^{(1)}_{3/2}(r_W/a)
\end{equation}
where $m_W$ and $R_W$ are the mass and radius of the wire, and $b^{(1)}_{3/2}$ is the standard Laplace coefficient. Note that this expression does not involve any individual harmonic coefficients, so we do not need to include a large series of terms in our calculations of the precession rate, or worry about prematurely truncating the series. 

Combining the Maclaurin spheroid with the equatorial wire yields an easily-computed model of Saturn's gravitational field with only 5 parameters: $J_2$, $J_4$, $J_6$ (or $\ell$), $m_W$ and $R_W$.  The parameters $J_2$ and $J_4$ are both well-constrained by existing observations and so can be treated as fixed for the purposes of the present analysis. However, when computing precession rates, one must account for the fact that the wire mass contributes to both $J_2$ and $J_4$:
\begin{equation}
J_{i,wire}=\frac{m_W}{M_S}\left(\frac{R_W}{R_S}\right)^iP_i(0)
\end{equation}
Thus, these terms should be removed from the estimates of $J_2$ and $J_4$ before
computing the harmonic coefficients and precession rate due to the pseudo-spheroid. 
Assuming this is done properly, then the remaining free parameters are $J_6$ (a measure of the contributions from Saturn's rotation-induced oblateness), and $m_W$ and $R_W$ (measures of  the contribution from Saturn's equatorial jet). In practice, we find that $m_W$ and $R_W$ are effectively degenerate with each other, and so we may further simplify the model by assuming $R_W=R_S$. This leaves a two-dimensional parameter space, which is much easier to visualize.  Of course, we caution the reader against over-interpreting these parameters. For example, $m_W$ does not necessarily equal the mass of Saturn's equatorial jet (note this simplistic model does not require the gravity and buoyancy of the jet to be properly balanced). Instead, it is probably best to think of $J_6$ and $m_W$ as measures of the relative contributions of the lower-order and higher-order gravitational harmonics, respectively.

\begin{figure}
\resizebox{6in}{!}{\includegraphics{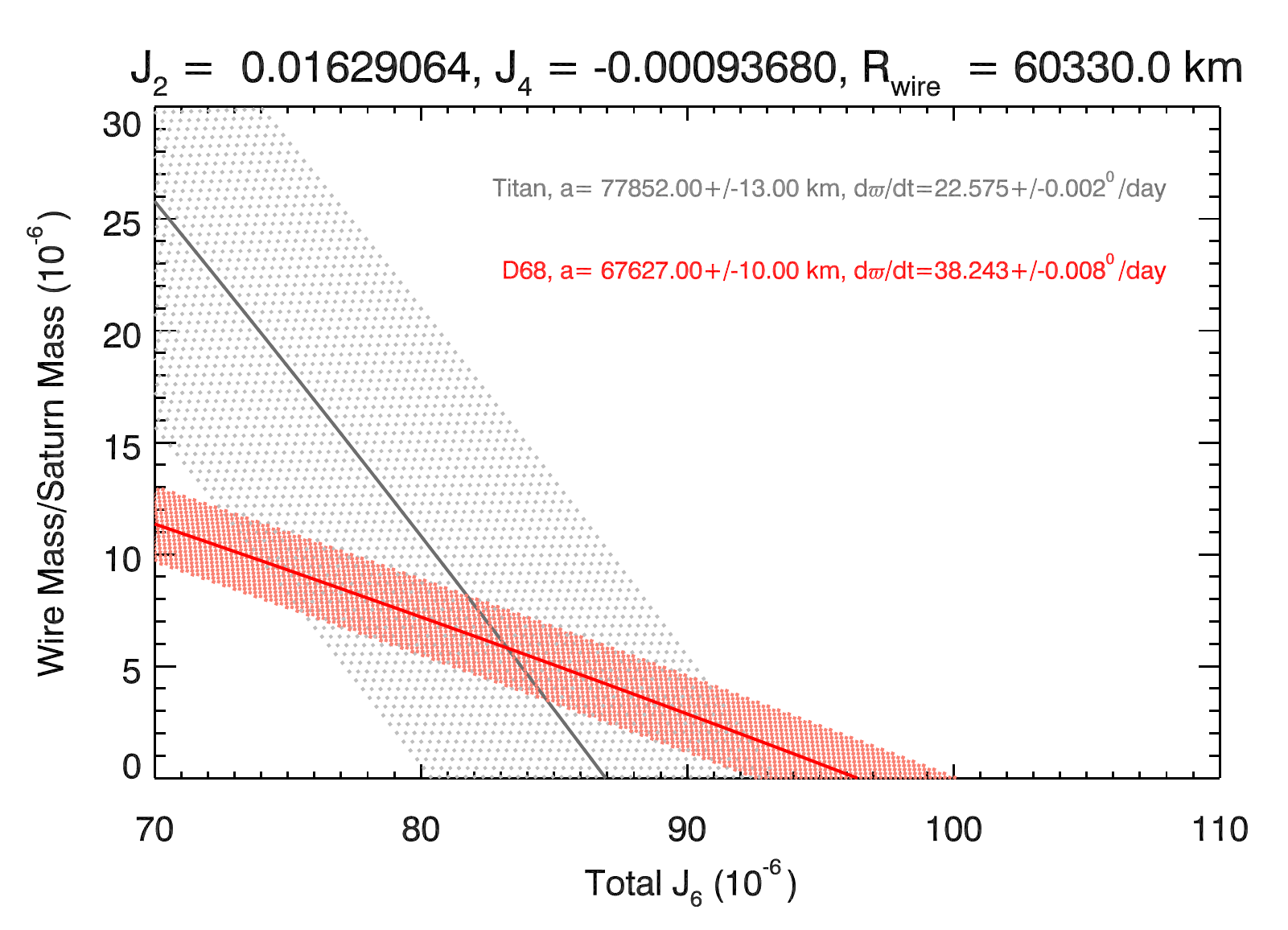}}
\caption{Plot showing the regions of $J_6-m_W$ space consistent with the observed precession rates of the Titan ringlet \citep{NP88} and D68 (this work), assuming the best-fit values of $J_2$ and $J_4$ without the Titan ringlet constraint \citep{Jacobson06}. Note that D68, being closer to the planet, is more sensitive to the higher-order gravity moments, and the resulting acceptance region has a different slope. Both ringlets' precession rates are consistent with a $J_6$ around $85\times10^{-6}$, which matches the estimate from \citet{Jacobson06} based on the motions of Saturn's moons.}
\label{conplot}
\end{figure}

Figure~\ref{conplot} shows the regions of $J_6-m_W$ parameter space that are consistent with the observed precession rates of D68 and the Titan ringlet \citep{NP88}. Note that the precession rate for each feature selects out a diagonal band in this space. This is sensible because the precession rate is a sum of the contributions from the rotating planet and the winds, and these contributions can be traded off against each other. At the same time, the bands derived from different features have different slopes. This is because features closer to the planet are more sensitive to higher-order moments, and so are more sensitive to changes in $m_W$ than $J_6$. Hence there is only a small region of parameter space that is consistent with the measured precession rates of both features. Note that this combined solution has a $J_6$ of around $85\times10^{-6}$, which is consistent with other estimates based strictly on Saturn's moons, and thus is insensitive to the assumed value of $m_W$ \citep{Jacobson06}. Also, the best-fit solution for $m_W$ is slightly positive, which would be consistent with a non-trivial anomaly due to the planet's equatorial jet. These results give us some hope that this parameterization of the planet's gravitational field has some value, but again we urge the reader not to take these parameters too literally. With only two observables and two unknowns, we were bound to find a solution somewhere.

The real value of this formalism will come in the future, when we can include data from additional non-circular features at different radii. If multiple measurements of precession rates all are consistent with the same values of $m_W$ and $J_6$, then we can have some confidence that this simple model provides a useful approximation of Saturn's gravitational field. On the other hand, if different constraints select out different regions of this parameter space, then this could be evidence that a more sophisticated model of Saturn's gravitational field is needed. Alternatively, this could reveal other perturbing forces that could be influencing precession rates for particular ringlets.

\section{Summary }
\label{summary}

The above analysis of various measurements of D68's brightness and position in Cassini images reveals the structure and dynamics of this ringlet are more complex than one might have expected. In particular:
\begin{itemize}
\item D68 exhibits longitudinal brightness variations that rotate around the planet at around 1751.65$^\circ$/day, and that might also evolve over time-scales of years.
\item D68 possesses a substantial eccentricity  ($ae =25\pm1$ km) and precesses around the planet at a rate of $38.243\pm0.008^\circ$/day, which is roughly consistent with current models of Saturn's gravity field.
\item D68's radial position does not exhibit obvious variations that can be attributed to normal mode oscillations with $m\ne1$.
\item D68's mean radius decreased at a rate of $2.4\pm0.4$ km/year between 2005 and 2013. D68 also seems to have moved outward between the Voyager and Cassini epochs, and the Cassini data hint that its mean radius may move back and forth with a period of around 15 Earth years.
\end{itemize}
The processes responsible for these phenomena and their potential implications for the Saturn system are still unclear. Nevertheless, D68's proximity to the planet means that the precession and migration rates derived here should provide important constraints on the gravitational and electromagnetic environment in the vicinity of Saturn.

\section*{Acknowledgements}

We wish to thank the Cassini Imaging Team, the Cassini Mission and NASA for providing the images that made this investigation possible. This work was supported by NASA Cassini Data Analysis Program Grant NNX12AC29G. We also wish to thank M.S. Tiscareno and M.W. Evans for their help with some of the image reduction procedures, as well as P.D. Nicholson and D. Hamilton for useful conversations. Finally, we would like to thank two reviewers for their helpful comments on this manuscript.


\end{document}